\documentclass[11pt]{article}

\usepackage[final]{acl}

\usepackage[T1]{fontenc}
\usepackage[utf8]{inputenc}
\usepackage{times}
\usepackage{latexsym}
\usepackage{microtype}
\usepackage{amsmath}
\usepackage{amssymb}

\usepackage{enumitem}
\usepackage{xcolor}
\newcommand{\zv}{\textcolor{red!70!black}{0}}           %
\newcommand{\best}[1]{\textcolor{green!50!black}{\bf #1}} %
\usepackage{xspace}

\usepackage{booktabs}
\usepackage{listings}
\usepackage{array}
\usepackage{multirow}
\usepackage{longtable}
\usepackage{cleveref}
\usepackage{tcolorbox}
\usepackage{subcaption}
\usepackage[font=small]{caption}
\usepackage{placeins}
\usepackage[normalem]{ulem} %
\usepackage{pifont} %
\usepackage{colortbl} %

\definecolor{yamlkeycolor}{RGB}{18,123,169}    %
\definecolor{yamlstringcolor}{RGB}{80,143,79}  %
\definecolor{yamlcommentcolor}{RGB}{128,128,128} %
\definecolor{yamlbackground}{RGB}{250,250,250} %
\definecolor{yamlframe}{RGB}{220,220,220}      %

\lstdefinestyle{yamlstyle}{
    basicstyle=\ttfamily\footnotesize,
    backgroundcolor=\color{yamlbackground},
    breakatwhitespace=true,
    breaklines=true,
    captionpos=b,
    commentstyle=\color{yamlcommentcolor},
    deletekeywords={...},
    escapeinside={\%*}{*)},
    extendedchars=true,
    frame=single,
    framesep=2pt,
    keepspaces=true,
    keywordstyle=\color{yamlkeycolor}\bfseries,
    morekeywords={type, output, schema, prompt},
    numbers=left,
    numbersep=5pt,
    numberstyle=\tiny\color{yamlcommentcolor},
    rulecolor=\color{yamlframe},
    showspaces=false,
    showstringspaces=false,
    showtabs=false,
    stepnumber=1,
    stringstyle=\color{yamlstringcolor},
    tabsize=1,
    xleftmargin=10pt,
    xrightmargin=2pt,
    framexleftmargin=10pt,
    framexrightmargin=2pt,
    framexbottommargin=2pt,
    framextopmargin=2pt,
    columns=flexible,
    linewidth=\columnwidth,
    basewidth=0.3em,
    lineskip=0.5pt,
    literate=
     *{:}{{\textcolor{yamlkeycolor}{:}}}{1}
     {|}{\textcolor{yamlkeycolor}{|}}{1}
     {\ \ }{{\ \ }}{1},
    emph={type,output,schema,prompt,validate,num\_retries\_on\_validate\_failure,operations,pipelines,input,model,join\_key,comparison\_model,comparison\_prompt,reduce\_key,resolution\_prompt,resolution\_model,output\_keys,prompts,content\_key,peripheral\_chunks,split\_key,chunk\_size,main\_chunk\_start,main\_chunk\_end,gleaning,unnest\_key,expand\_fields,datasets},
    emphstyle={\color{yamlkeycolor}\bfseries}
}

\lstnewenvironment{yaml}[1][]
  {\lstset{style=yamlstyle, #1}}
  {}

\lstdefinestyle{plaintextstyle}{
  breaklines=true,
  breakatwhitespace=false,
  basicstyle=\footnotesize\ttfamily,
  columns=flexible,
  keepspaces=true,
  showstringspaces=false,
  frame=single,
  framesep=2pt,
  numbers=left,
  numbersep=5pt,
  numberstyle=\tiny\color{gray},
  xleftmargin=10pt,  %
  framexleftmargin=10pt,
  xrightmargin=3pt,
  resetmargins=true, %
  aboveskip=10pt,    %
  belowskip=10pt,    %
  gobble=0           %
}

\lstdefinelanguage{text}{
  identifierstyle=,
  keywordstyle=,
  commentstyle=,
  stringstyle=,
  morekeywords={}
}

\newcommand{\techreport}[1]{#1}
\newcommand{\hidecr}[1]{}
\newcommand{\papertext}[1]{}

\usepackage{tcolorbox}
\newtcbox{\fmbadge}{
  on line, arc=2pt, outer arc=2pt,
  colback=violet!8, colframe=violet!35,
  boxsep=0pt, left=3pt, right=3pt, top=1pt, bottom=1pt,
  boxrule=0.4pt,
  fontupper=\small\sffamily\bfseries
}
\newtcbox{\smallfmbadge}{
  on line, arc=1.5pt, outer arc=1.5pt,
  colback=violet!8, colframe=violet!35,
  boxsep=0pt, left=2pt, right=2pt, top=0pt, bottom=0pt,
  boxrule=0.3pt,
  fontupper=\scriptsize\sffamily\bfseries
}
\newcommand{\ds}[1]{{\textit{\textcolor{gray!70!black}{\MakeLowercase{#1}}}}}
\newcommand{\dst}[1]{{\scriptsize \textit{\textcolor{gray!70!black}{\MakeLowercase{#1}}}}}

\newcommand{\del}[1]{}
\newcommand{\cdel}[1]{}
\newcommand{\cnew}[1]{\textcolor{black}{\relax#1}}

\newcommand{\crdy}[1]{{\color{black}#1}}

\newcommand{\blank}[1]{\textcolor{gray!60!white}{\em #1}}

\newcommand{\takeaway}[1]{%
  \vspace{2pt}\par\noindent
  \centerline{\colorbox{blue!8}{\parbox{\dimexpr\columnwidth-2\fboxsep}{%
    {\scriptsize\textsf{\textbf{Takeaway:}}} \footnotesize #1.}}}%
  \vspace{1pt}\par\noindent}

\newcommand{\topic}[1]{\vspace{-3.5pt}\smallskip \smallskip \noindent{\bf #1.}}
\newcommand{\bench}{\textsc{DAB}\xspace}
\newcommand{\benchurl}{\techreport{\href{https://github.com/ucbepic/DataAgentBench}{\ttt{github.com/ucbepic/DataAgentBench}}}\papertext{{\color{red!70!black}[URL redacted for submission]}}}
\newcommand{\ttt}[1]{{\small \texttt{#1}}\xspace}
\newcommand{\tinytt}[1]{{\scriptsize \texttt{#1}}\xspace}

\newcommand{\greencheck}{\textcolor{green}{\ding{52}}} %
\newcommand{\redcross}{\textcolor{red}{\ding{55}}}     %
\newcommand{\partialcheck}{{\color{orange}$\circ$}}  %

\begin{document}
\title{Can AI Agents Answer Your Data Questions?\\A Benchmark for Data Agents}

\author{
  Ruiying Ma\textsuperscript{1}\thanks{\hspace{2pt}Co-first authors.},
  Shreya Shankar\textsuperscript{1}\footnotemark[1],
  Ruiqi Chen\textsuperscript{2},
  Adit Chintamaneni\textsuperscript{1},\\
  \bfseries Yiming Lin\textsuperscript{1},
  Sepanta Zeighami\textsuperscript{1}, 
  Rajoshi Ghosh\textsuperscript{3},
  Abhinav Gupta\textsuperscript{3},\\
  \bfseries Anushrut Gupta\textsuperscript{3},
  Tanmai Gopal\textsuperscript{3},
  Aditya G. Parameswaran\textsuperscript{1} \\[4pt]
  \textsuperscript{1}UC Berkeley,
  \textsuperscript{2}University of Washington,
  \textsuperscript{3}Hasura PromptQL \\
  \texttt{\{ruiyingm, shreyashankar, adityagp\}@berkeley.edu}
}

\maketitle

\begin{abstract}
Users across enterprises increasingly rely on LLM agents to query their
data through natural language. However, building reliable {\em data agents}
remains difficult because real enterprise data is fragmented across many
heterogeneous database systems, with duplicated and inconsistent data,
and key information often buried in unstructured text---requiring agents
to go beyond just writing SQL or data science scripts to answer
questions. 
Existing benchmarks tackle only individual pieces of this
end-to-end workflow---e.g., text-to-SQL over a single database---and are
increasingly saturated and contaminated. 
We present a new benchmark for LLM agents,
the Data Agent Benchmark (DAB), grounded in a study of enterprises
building production data agents across six industries. 
DAB comprises 104
queries across 17 datasets and 4 database management systems. On DAB,
the best state-of-the-art agent achieves only 57\% pass@1. 
We analyze
agent failure modes and distill takeaways for future data-agent
development. Our benchmark and experiment code are published at \benchurl.
\end{abstract}

\begin{figure*}[t]
    \centering
    \begin{minipage}{0.75\textwidth}
    \centering
    \begin{subfigure}[b]{0.26\textwidth}
        \centering
        \includegraphics[width=\linewidth]{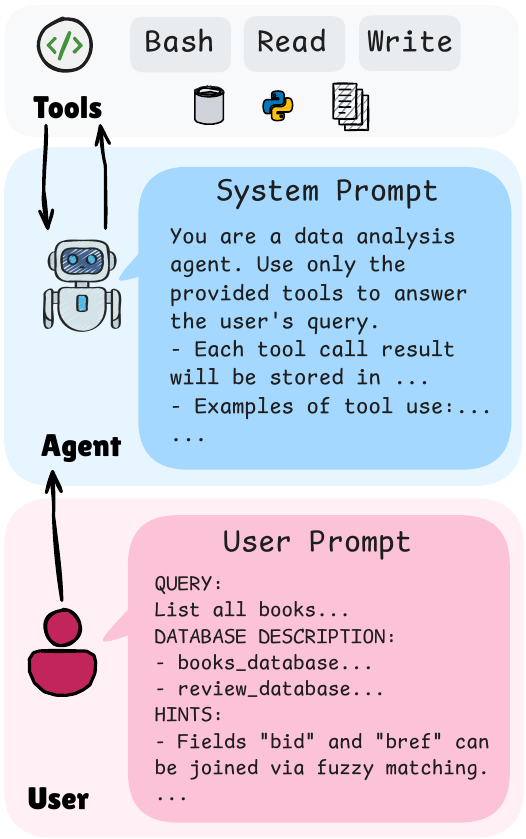}
        \caption{\bench setup.}
    \end{subfigure}
    \hfill
    \begin{subfigure}[b]{0.7\textwidth}
        \centering
        \includegraphics[width=\linewidth]{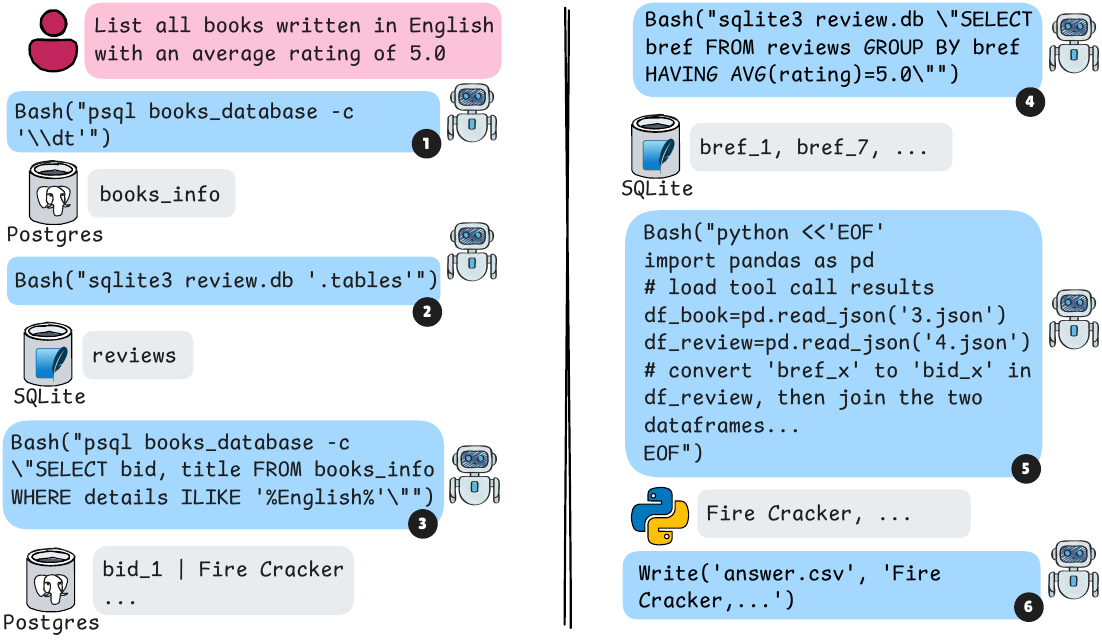}
        \caption{Example agent trace.}
    \end{subfigure}
    \end{minipage}
    \caption{(a)~In \bench, an agent answers a user query by
interacting with tools provided by its harness (e.g., Claude Code, the OpenAI Codex SDK).
(b)~The agent extracts information
embedded in a free-text column (\ttt{details}) via SQL,
integrates data across PostgreSQL and SQLite, and
reconciles ill-formatted join keys (\ttt{bref} vs.\
\ttt{bid}).}
    \label{fig:agent-setup}
\vspace{-10pt}
\end{figure*}

\section{Introduction}
\label{sec:intro}

Users across enterprises increasingly want {\em data agents}, or
AI agents that answer natural-language questions
over their data.
Data agents are a major application of LLMs and a new
workload pattern for data systems~\cite{liu2026agentfirst}:
major database vendors have added agent capabilities to
their platforms~\cite{snowflake2025cortex, databricks2025assistant},
and organizations have invested heavily in building their
own---for example,
Uber's QueryGPT handles over 1.2 million interactive queries
per month~\cite{uber2024querygpt},
and OpenAI built an internal data agent
used by thousands of employees
to query 70,000 datasets
totaling 600~petabytes~\cite{xu2026openai}.

Yet building reliable data agents is hard.
Consider an example query for a data agent:
a sales analyst asks,
{\em ``Which leads from last quarter should we follow up on?''}
Answering this requires
finding lead records in a customer relationship management (CRM) tool,
matching them against call transcripts
stored in a separate document database,
classifying each lead's intent
from unstructured text,
and applying domain knowledge
about what makes a lead good to pursue.
In fact, enterprise data is well-known to be
fragmented and messy: surveys find that
72\% of organizations store data
in disparate silos~\cite{stitchdata2024silos}
and 82\% report that these silos
disrupt critical workflows~\cite{ibm2024datadiff}.

Unfortunately, no benchmark measures
{\em end-to-end} data agent capabilities.
For example, text-to-SQL benchmarks~\cite{yu2018spider, li2023can, leispider, chen2025beaver}
test whether LLMs can translate
a natural-language question
into a single correct query
over a single relational database,
but do not require multi-step reasoning
or integration across different databases.
Table question-answering (Table-QA)
benchmarks~\cite{chen2020hybridqa, chen2021finqa}
provide tiny tables entirely in the prompt---unlike
enterprise workloads, where tables live in databases
and rarely fit in context.
Moreover, the public benchmarks in this space are
increasingly saturated by frontier models, and reported
scores are further confounded by two issues: data
contamination of public test sets in
pretraining~\cite{ranaldi2024contamination}, and pervasive
annotation errors that inflate accuracy---error rates
reach 52.8\% in BIRD Mini-Dev and 62.8\% in
Spider-Snow~\cite{jin2026pervasive}.

\topic{The Data Agent Benchmark (\bench)} We present
\bench, a new benchmark for
evaluating AI agents on realistic, complex data-oriented
tasks. {\bf {\em \bench comprises 104 natural-language
queries across 17 datasets and 4 database management
systems (DBMSes).}} The size of \bench is comparable to
other carefully-curated agent benchmarks
(e.g., FrontierMath~\cite{glazer2024frontiermath},
TerminalBench~\cite{merrill2026terminal}).
\crdy{Each \bench query requires an agent to interact with a full environment using database and Python tools, potentially involving hundreds of reasoning and action steps. This differs substantially from large-scale benchmarks such as Text-to-SQL, where each item typically requires a single LLM call (e.g., to generate one SQL query).}
\bench has been publicly available\techreport{\footnote{\url{https://ucbepic.github.io/DataAgentBench/}}}\papertext{\footnote{{\color{red!70!black}[URL redacted for submission]}}}
for \cdel{two}\cnew{five} months and has already attracted \cdel{17}\cnew{29} submissions.

Building \bench took us over a year and involved three
core challenges.
The first challenge was to solicit aspects of production
workloads that make data-agent queries difficult. Through
a formative study with six enterprises building production
data agents (\Cref{sec:bench-src-char}), we identified
four properties that are unaddressed by existing benchmarks:
{\em (i)}~the data needed to answer a query is distributed
across multiple heterogeneous database systems,
{\em (ii)}~identifiers for the same entity are formatted
differently across databases,
{\em (iii)}~answers are buried in unstructured text
fields and cannot be retrieved via SQL, and
{\em (iv)}~the query requires domain knowledge not
inferable from the data alone (e.g., finance jargon).

The second challenge was to translate these properties
into a reproducible benchmark. Since enterprise data is
proprietary, we had to build \bench from open-source
data---which is typically clean---to mirror the dirty,
duplicated, and inconsistently formatted data we observed
in production. 
We also had to ensure there is a deterministic, unambiguous ground-truth answer for each query.

The third challenge was calibrating difficulty: \bench
should be hard enough to discriminate among agents, but
not so hard that no agent succeeds. We spent \$4{,}346 to evaluate three
state-of-the-art frontier agents and harnesses---Claude-Sonnet-4.6
(Claude Code), GPT-5.5 (Codex), and Gemini-3.1-Pro
(ReAct)---each under three experimental variants: no
assistance, with database and schema descriptions added,
and with both descriptions and hints describing how the
data was corrupted. The best agent reaches only
\textbf{{\em 57\% pass@1}}
(5 trials per query),
and most datasets show partial accuracy across agents,
leaving headroom for improvement. We further analyze
trajectories and provide a taxonomy of failure modes.
90\% of wrong answers stem from incorrect
planning or faulty implementation rather than wrong data
selection; and agents rely almost exclusively on regular
expressions for free-text extraction, with only one agent
attempting alternatives like using NLP parsing libraries
or spawning subagents to extract values in context.

In summary, this paper makes the following contributions:
\begin{enumerate}[leftmargin=0pt, itemindent=*,
                  labelsep=4pt, widest=3, align=left, nosep]
    \item We characterize real-world data-agent workloads
    via a study with six enterprises building data agents,
    and identify four properties that distinguish them
    from the queries existing benchmarks cover.

    \item We construct and release \bench, a new benchmark
    of 104 queries across 17 datasets and 4 database
    systems.

    \item We evaluate state-of-the-art agents on \bench
    and find that even the best agent achieves only 57\%
    pass@1;
    our error
    analysis surfaces failure modes that point to concrete
    directions for improving data agents.
\end{enumerate}

\section{Related Work}
\label{sec:related}

Despite databases being among the most ubiquitous
professional tools, no frontier-model evaluation
includes a database-use benchmark.
We position \bench against a few categories of benchmarks\techreport{, summarized in \Cref{tab:comp-bench}}.

\topic{Tool-use benchmarks}
Tool-use benchmarks evaluate LLM agents on external tools
such as function calling~\cite{patilberkeley}, code
editing~\cite{jimenez2024swe}, browser
interaction~\cite{zhou2024webarena}, command-line
tasks~\cite{merrill2026terminal}, and multi-step web
reasoning~\cite{mialon2023gaia}, but none involves querying
a database. Closer to our setting, benchmarks for
data-adjacent workflows such as retail customer
service~\cite{yao2025taubench} and CRM
operations~\cite{huang2025crmarena} expose each data
operation through a hand-crafted API endpoint---the agent
never queries the underlying database directly.

\topic{Data-related benchmarks}
Other benchmarks target data tasks directly, yet each
captures only part of the data-agent workflow.

\textit{Table question-answering.}
Table QA benchmarks supply data in the prompt and ask the
model to reason over in-context
tables~\cite{pasupat2015compositional, chen2020hybridqa,
talmor2021multimodalqa, chen2021finqa, zhu2021tatqa};
some add unstructured text, as in
HybridQA~\cite{chen2020hybridqa}, or domain knowledge, as
in FinQA~\cite{chen2021finqa}.
Because tables are given in context, the agent never
queries a database, and real-world tables routinely exceed
context-window limits.

\textit{Text-to-SQL.}
Text-to-SQL benchmarks instead have the model generate a
query for a natural-language question over a single
database---Spider~\cite{yu2018spider},
WikiSQL~\cite{zhong2017seq2sql}, and BIRD~\cite{li2023can},
with extensions to multi-turn dialogue~\cite{yu2019cosql}
and cloud SQL dialects~\cite{leispider}.
This setup is limited: it tests only SQL {\em generation}
over a single clean database, whereas enterprise questions
typically span multiple systems and require transformations
that SQL cannot express.
Even within this narrow setting, text-to-SQL performance on
data-warehouse tasks over private data is
abysmal~\cite{chen2025beaver}. These text-to-SQL benchmarks
are also saturated and corrupted:
top systems reach 91.2\% execution accuracy on
Spider~\cite{yu2018spider}, and the reported scores are
confounded by data contamination---the public test sets
appear in pretraining corpora~\cite{ranaldi2024contamination}---and
by annotation errors, with error rates of 52.8\% in BIRD
Mini-Dev and 62.8\% in Spider~2.0-Snow~\cite{jin2026pervasive}.

\textit{Data science and engineering.}
Beyond SQL, other benchmarks task the agent with analyzing
data, but over flat files (e.g., CSVs) rather than
databases. DS-1000~\cite{lai2023ds},
DA-Code~\cite{huang2024code}, and
KramaBench~\cite{lai2025kramabench} require multi-step code
over such files, while Spider~2-V~\cite{cao2024spider2}
requires operating data-engineering tools through GUI
actions (e.g., configuring an Airbyte sync).
Neither setting requires querying a database directly.

\techreport{%
\textit{Semantic query processing.}
Semantic query processing extends relational operators with
natural-language variants powered by
LLMs~\cite{jo2024thalamusdb, liu2025palimpzest,
patel2025semantic, shankar2025docetl, liskowski2025cortex}.
Benchmarks such as SemBench~\cite{lao2025sembench} and
TAG-Bench~\cite{biswal2024text2sql} execute a
pre-constructed pipeline of these operators over flat files
to obtain a single answer, but do not capture multi-system
integration. The semantic operations in \bench are a subset
of those these benchmarks evaluate, embedded within an
end-to-end querying workflow over real database systems.%
}

Overall, each of these benchmarks covers only a slice of data
work, and many prescribe the exact tools or operations the
agent must use. \bench takes a different view: it
specifies a realistic data setup---multiple databases with
messiness seen in production workloads---and places no
restrictions on how the agent solves the task.

\techreport{%
\begin{table}[t]
    \centering
    \small
    \setlength{\tabcolsep}{4pt}
    \renewcommand{\arraystretch}{0.9}
    \caption{Coverage of \bench's four properties:
    (i)~multi-database integration,
    (ii)~ill-formatted join keys,
    (iii)~unstructured text transformation,
    (iv)~domain knowledge.
    \greencheck~yes, \partialcheck~partial, \redcross~no.}
    \label{tab:comp-bench}
    \begin{tabular}{@{}lcccc@{}}
    \toprule
    {\bf Benchmark}& {\bf (i)} & {\bf (ii)} & {\bf (iii)} & {\bf (iv)} \\
    \midrule
    General tool-use~\cite{patilberkeley, jimenez2024swe} & \redcross & \redcross & \redcross & \partialcheck \\
    Table QA~\cite{chen2020hybridqa, chen2021finqa} & \redcross & \redcross & \greencheck & \partialcheck \\
    Text-to-SQL~\cite{leispider, li2023can} & \redcross & \redcross & \redcross & \partialcheck \\
    Data science \& eng.~\cite{huang2024code, lai2025kramabench} & \redcross & \redcross & \redcross & \redcross \\
    Semantic query processing & \redcross & \redcross & \greencheck & \redcross \\
    \bench & \greencheck & \greencheck & \greencheck & \greencheck \\
    \bottomrule
    \end{tabular}
    \vspace{-10pt}
\end{table}%
}

\begin{table*}[t]
    \centering
    \caption{Overview of datasets and queries in \bench. DBMS: M=MongoDB, P=PostgreSQL, S=SQLite, D=DuckDB. \#Tbl=tables, \#Q=queries, \#DK=queries requiring domain knowledge (property~(iv)).}
    \label{tab:bench-overview}
    \scriptsize
    \setlength{\tabcolsep}{2.5pt}
    \renewcommand{\arraystretch}{1.15}
    \begin{tabular}{@{}>{\raggedright\arraybackslash}p{0.10\textwidth}rlrrr p{0.66\textwidth}@{}}
        \toprule
        \textbf{Dataset} & \textbf{\#DBs} & \textbf{DBMS} & \textbf{\#Tbl} & \textbf{\#Q} & \textbf{\#DK} & \textbf{Example Query} \\
        \midrule
        \dst{agnews} & 2 & M,S & 3 & 4 & 0 & {\em What is the title of the sports article whose description has the greatest number of characters?} \\[1pt]
        \dst{bookreview} & 2 & P,S & 2 & 3 & 0 & {\em Which English-language books in the `Literature \& Fiction' category have a perfect average rating of 5.0?} \\[1pt]
        \dst{civic\_unstructured} & 2 & M,S & 2 & 10 & 5 & {\em As of 2023-01-01, list all projects whose most recent report records both a type of `capital' and a status of `design' and that have accumulated total funding of more than \$500{,}000.} \\[1pt]
        \dst{crmarenapro} & 6 & D,P,S & 27 & 13 & 10 & {\em Which states have the quickest case closure time in the past 6 quarters?} (Assume today's date is 2022-10-26) \\[1pt]
        \dst{cve} & 4 & D,M,P,S & 6 & 10 & 10 & {\em How many distinct CVEs published in 2023 are in CISA's KEV catalog AND have at least one vulnerable Apache product configuration?} \\[1pt]
        \dst{deps\_dev\_v1} & 2 & D,S & 3 & 2 & 2 & {\em Among all NPM packages with license `MIT' marked as release, which 5 have the highest GitHub fork count?} \\[1pt]
        \dst{github\_repos} & 2 & D,S & 6 & 4 & 4 & {\em Among repos not using Python, what proportion of README.md files include copyright information?} \\[1pt]
        \dst{googlelocal} & 2 & P,S & 2 & 4 & 0 & {\em What are the top 5 businesses in Los Angeles, CA, ranked by highest average rating?} \\[1pt]
        \dst{imdb} & 2 & P,S & 6 & 10 & 3 & {\em Among titles produced after 1990 with a US release date in the 1990s--2000s and a verified cast listing, what is the alphabetically first title?} \\[1pt]
        \dst{krama} & 3 & M,S & 4 & 10 & 8 & {\em Among NIFC regions with a Geographic Area Coordination Center (GACC), which GACC accumulated the most human-caused wildfire acres from 2020--2023?} \\[1pt]
        \dst{music\_brainz\_20k} & 2 & D,S & 2 & 3 & 0 & {\em Which store earned the most USD revenue from Brucqe Maginnis' song `Street Hype' across all countries?} \\[1pt]
        \dst{pancancer\_atlas} & 2 & D,P & 3 & 3 & 3 & {\em Among alive BRCA patients, which top 3 histological types show the highest \% of CDH1 gene mutations?} \\[1pt]
        \dst{patents} & 2 & P,S & 2 & 3 & 3 & {\em Identify CPC areas with the highest EMA of patent filings (smoothing 0.2); return level-5 codes whose best year is 2022.} \\[1pt]
        \dst{stockindex} & 2 & D,S & 2 & 3 & 3 & {\em Which stock index in Asia has the highest average intraday volatility since 2020?} \\[1pt]
        \dst{stockmarket} & 2 & D,S & 2754 & 5 & 5 & {\em List all ETFs on NYSE Arca with adjusted close above \$200 at any point in 2015; report the total count.} \\[1pt]
        \dst{usaspending} & 4 & D,M,P,S & 6 & 10 & 8 & {\em How many distinct Department of Defense contract awards have an amount greater than \$1{,}000{,}000?} \\[1pt]
        \dst{yelp} & 2 & D,M & 5 & 7 & 0 & {\em During 2018, how many reviewed businesses offered either business parking or bike parking?} \\
        \bottomrule
    \end{tabular}
    \vspace{-10pt}
\end{table*}

\begin{figure}[t]
    \centering
    \includegraphics[width=\linewidth]{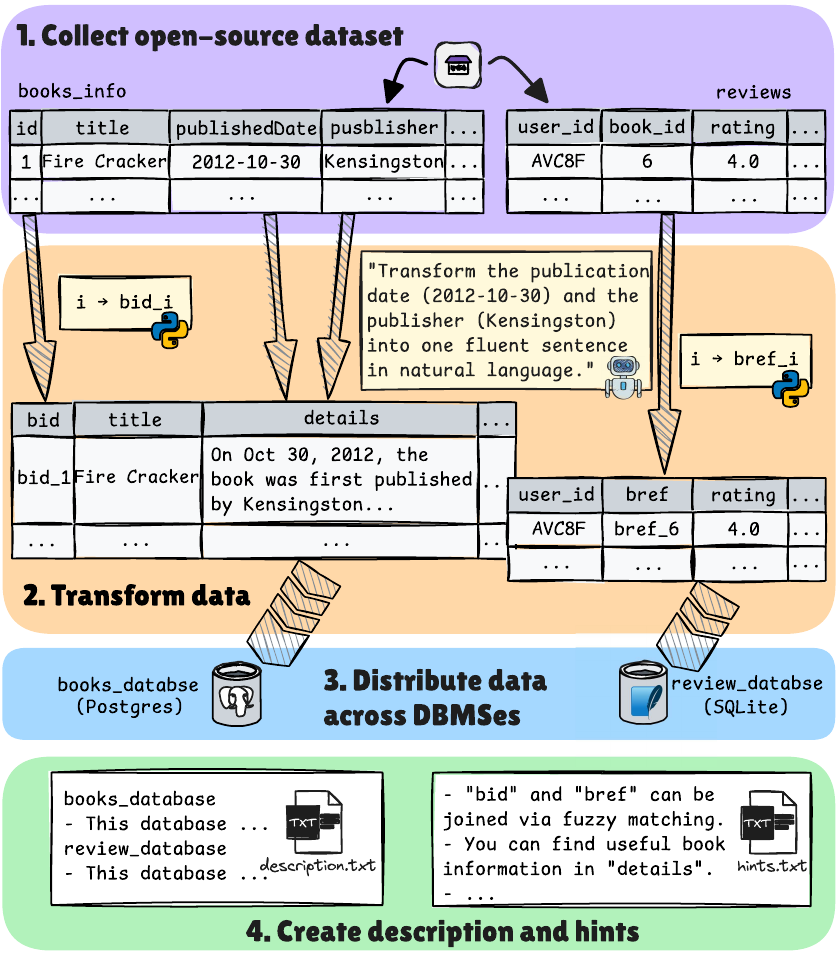}
    \caption{Dataset creation methodology, illustrated on the \ds{bookreview} dataset.
    Step~1: collect an open-source dataset with two tables, \ttt{books\_info} and \ttt{reviews}.
    Step~2: transform the data by removing \ttt{publishedDate} and \ttt{publisher} and re-embedding their values into a new \ttt{details} column via LLM-generated sentences, and by prefixing the join keys (\ttt{id}~$\to$~\ttt{bid}, \ttt{book\_id}~$\to$~\ttt{bref}).
    Step~3: distribute the tables across PostgreSQL and SQLite.
    Step~4: create a dataset description (\ttt{descriptions.txt}) and a hints file (\ttt{hints.txt}).}
    \vspace{-20pt}
    \label{fig:dataset-creation}
\end{figure}

\section{Benchmark Construction}
\label{sec:bench}

\subsection{Formative Study}
\label{sec:bench-src-char}

We conducted a study (approximately two months of
interviews and analysis) with six enterprises building
production data agents across six industries (technology,
finance, food services, e-commerce, SaaS, and healthcare).
The interview protocol and an anonymized summary of one
enterprise's response are in
\Cref{append-sec:interview-protocol}.
We performed a thematic analysis~\cite{clarke2017thematic}
of the responses to find recurring patterns of challenges,
and came up with the following four properties of the
workloads that should be in \bench:

\begin{enumerate}[label={\em (\roman*)}, leftmargin=0pt, itemindent=*,
                  labelsep=4pt, widest=iii, align=left, nosep]
\item \textbf{multi-database integration}: a single question may require querying across several databases with different query language dialects (e.g., SQL dialects and MongoDB's query language).
\item \textbf{ill-formatted join keys}: identifiers for the same entity may be formatted differently across databases---e.g., a book ID appearing as \ttt{bid\_123} in one table and \ttt{bref\_123} in another---requiring the agent to detect and reconcile the mismatch before joining.
\item \textbf{unstructured text transformation}: answers may be embedded in text fields that the agent must parse into structured values before they can be filtered, grouped, or joined.
\item \textbf{domain knowledge}: answering the query correctly requires expertise not inferable from the data alone, such as knowing that stock volatility must be computed from adjusted closing prices to account for splits and dividends.
\end{enumerate}

We exclude two common types of queries: those that depend
on live external APIs, and open-ended queries (e.g.,
``why are my sales going down?''), since neither admits a
single stable ground-truth answer.

\subsection{Construction}
\label{sec:bench-constr-method}

\Cref{tab:bench-overview} details the 17 datasets in
\bench with a representative query each.
Because the enterprises' data is proprietary, we build
\bench from open-source datasets whose queries \del{mirror}\crdy{instantiate} the
four properties identified above. Every query in \bench
involves~(i) and at least one of (ii) or (iii); (iv)
appears in proportion to its prevalence in the study.
Constructing \bench took approximately eight months of
author time, averaging roughly 50~hours per dataset to
collect the source data, design and corrupt it, write the
queries, and validate the ground-truth answers.

\topic{Datasets}
\Cref{fig:dataset-creation} illustrates the four-step
dataset creation procedure; we walk through them, using the \ds{bookreview} dataset as a running
example.

\begin{enumerate}[label=\textbf{(\arabic*)}, leftmargin=0pt, itemindent=*,
                  labelsep=4pt, widest=4, align=left, nosep]
\item \textbf{Collect open-source datasets.}
We collect open-source datasets that have multiple tables
(so we can later distribute them across databases) and
at least one unstructured text column (so we can later
embed answer-revealing values into it). Four of the 17 datasets---\ds{agnews}~\cite{zhang2015character},
\ds{crmarenapro}~\cite{huang2025crmarena}, \ds{imdb}~\cite{leis2015good}, and
\ds{krama}~\cite{lai2025kramabench}---come from existing
benchmarks; we sample their queries (rewriting some to
remove ambiguity). The remaining 14 are sourced from open
data repositories. For example, \ds{bookreview} consists
of a \ttt{books\_info} table (with an unstructured
\ttt{description} field) and a separate \ttt{reviews}
table.

\item \textbf{Induce ill-formatted join keys (property~ii)
and unstructured-text transformation (property~iii).}
We remove columns that would trivially answer a query and
re-embed their values elsewhere, so the
agent must work to recover them. Four example strategies include:
\begin{itemize}[leftmargin=0pt, itemindent=*, labelsep=4pt, align=left, nosep]
\item \emph{Renaming a shared identifier across tables.}
The same identifier appears under different column names
in the two tables it joins. For example, in
\ds{bookreview}, the book identifier is \ttt{book\_id} in
the \ttt{books\_info} table (PostgreSQL) but
\ttt{purchase\_id} in the \ttt{review} table (SQLite);
to join the two, the agent must recognize that these
columns refer to the same entity.

\item \emph{Applying many row-specific corruptions to a
column.}
For some datasets we apply one of several corruption
methods row-by-row to the same column, with so many
variants per row that no small regex can clean the
column: the agent has to inspect every row to figure out
the cleaning logic. The methods include:
\begin{itemize}[leftmargin=0pt, itemindent=*, labelsep=4pt, align=left, nosep]
\item[(a)] prepending a different domain-relevant word to
each value, generated by a coding agent (Claude Code)
that produces thousands of plausible words per dataset.
For \ds{usaspending}, these words resemble agency codes,
fiscal-year tags, or program identifiers (e.g.,
\ttt{FY2023\_}, \ttt{DOD-}, \ttt{PROG/}).
\item[(b)] applying per-character OCR-like substitutions
($0{\leftrightarrow}O$, $1{\leftrightarrow}I$,
$5{\leftrightarrow}S$, etc.) with some probability
(${\sim}0.1$ per character).
\end{itemize}
We apply these methods independently in each table where
the column appears. For example, in \ds{usaspending}, the
\ttt{award\_id} column is corrupted with OCR
substitutions, applied independently in the
\ttt{contracts} and \ttt{recipients} tables; to join the
two, the agent must essentially perform entity resolution
on the \ttt{award\_id} column.

\item \emph{Hiding answer-revealing values in a text
column.}
We prompt GPT-4o to rewrite a free-text column so it
mentions the value in prose (``transform \{free\_text\}
to naturally include a reference to \{value\}; change as
little as possible''), then drop the original column. For
example, in \ds{bookreview}, \ttt{publishedDate},
\ttt{publisher}, and \ttt{language} are embedded into
\ttt{details} and removed from \ttt{books\_info}.

\item \emph{Encoding a numeric column as varied free text.}
We convert a numeric column into a free-text column where
each row's value is written differently, requiring the
agent to do data cleaning to recover the underlying
number. For example, in \ds{usaspending}, the contract
award amount is stored as text where one row might read
\ttt{1500000}, another \ttt{\$1,500,000}, another
\ttt{1.5M}, another \ttt{one million five hundred
thousand dollars}, and another a sum-of-parts expression
like \ttt{\$1{,}050{,}000 + \$450{,}000}.
\end{itemize}

\item \textbf{Distribute tables across DBMSes (property~i:
multi-database integration).}
We split each dataset's tables across at least two DBMSes,
deciding placement manually to mirror how enterprises
typically organize data: transaction-oriented tables
(records, profiles, reviews) go in row-oriented or document
stores (PostgreSQL, MongoDB, SQLite), analytics-oriented
tables (large fact tables, time series) go in the columnar
store DuckDB, and large free-text or document corpora
(e.g., article bodies, review collections) go in MongoDB.
The agent must therefore handle MongoDB's query language
alongside SQL dialects. For example, \ds{bookreview}'s book
metadata table \ttt{books\_info} goes in PostgreSQL and its
\ttt{reviews} table in SQLite.

\item \textbf{Provide a natural-language description and
hints file (the latter inducing property~iv: domain
knowledge).}
The description lists each database's logical name and
schema; the hints file flags the transformations applied
and any domain-specific definitions. Both are shown in the
walk-through in \Cref{sec:bench-stats}.
\end{enumerate}

\topic{Queries and Answers}
\label{sec:task-formulation}%
\bench contains 104 queries, each incorporating at least two
of the four properties. A query comprises a
natural-language question, a ground-truth answer (derived
from the pre-transformation data), and a {\em validation
script}: a Python function that takes the agent's free-form
response\crdy{, normalizes it (e.g., trimming whitespace, lower-casing),} and returns True or False \del{using query-specific
logic}\crdy{ by checking whether the ground truth appears within the normalized string}. Validation scripts are bespoke per query\del{; for
example, checking substring containment for a single
value such as a book title, set membership for a list of
returned items, or numerical tolerance for a computed
metric.}\crdy{ according to the answer type as follow:
\begin{itemize}[leftmargin=0pt, itemindent=*, labelsep=4pt, align=left, nosep]
    \item Numeric answers: we check whether the response contains a value \cdel{falling }within a predefined tolerance range of the ground truth.
    \item Single-valued answers (e.g., a book title): the ground\cdel{-truth string}\cnew{ truth} \cdel{must }appear\cnew{s} as a substring of the response.
    \item Alternative valid formats: some answers\cdel{, while}\cnew{ are} semantically equivalent to the ground truth\cdel{, appear in alternative valid formats}\cnew{ but formatted differently}. For example, ETF names may appear with or without their share-class suffix (e.g., ``Vanguard S\&P 500 ETF'' vs. ``Vanguard S\&P 500 ETF VOO'')\cdel{, and repository names \cdel{may be written with or without }their owner prefix (e.g., ``pytorch/pytorch'' vs. ``pytorch''); since these formats differ from the ground truth}\cnew{. Since these differ} only slightly, we \cdel{instead }compute the Levenshtein distance between the response and the \cdel{answer}\cnew{ground truth} and accept it if \cdel{the distance is }below a threshold (on average, ~20\% of the \cdel{ground-truth }answer's length).
    \item Set-valued answers (e.g., a list of book titles): every element must appear, \cdel{with }each checked by the single-answer rule above\cdel{ (e.g., numeric or single-valued)}.
    \item Dictionary-valued answers (e.g., book titles with their average ratings): every key must appear, and each value {must appear within a surrounding window of} its key, with keys and values \cdel{each }checked by the single-answer rule above.
\end{itemize}
Our validation scripts check whether the response contains the complete correct answer, even if it also contains incorrect information. Detecting additional incorrect information would require manual review or an LLM judge, making evaluation not scalable or deterministic.} 
Two authors jointly write each query, its ground
truth, and its validation script.
To verify \bench, \del{one author manually inspected
every query, description, and hints file, and an industry
partner independently reran every query to confirm the
validation scripts.}
\crdy{we continue the following process until all reviewers reach consensus:
\begin{enumerate}[label={\em (\roman*)}, leftmargin=0pt, itemindent=*,
                  labelsep=4pt, widest=ii, align=left, nosep]
    \item Two authors independently review every \cdel{transformed }dataset, solve every query, and test every validation script, using \cdel{any }helpful tools (e.g., Claude for domain knowledge).
    \item They compare their findings and reconcile disagreements. We classify disagreements into the following two types and resolve \cdel{each according to its root cause}\cnew{by root cause}:
    \begin{itemize}[leftmargin=0pt, itemindent=*, labelsep=4pt, align=left, nosep]
    \item Dataset or query disagreements arise almost entirely from ambiguous \cdel{query }wording. We \cdel{iteratively }clarify the query and ask another author to \cdel{independently }solve each revision until the intended answer is unambiguous.
    \item Validation script disagreements arise when the script \cdel{does not account for all}\cnew{misses} valid answer formats or edge cases. We identify the missing cases and \cdel{iteratively }refine the script to handle them (e.g., \cdel{using }Levenshtein distance to \del{accommodate}tolerate minor string variations).
    \end{itemize}
\end{enumerate}
}

\subsection{Statistics and Walk-Through}
\label{sec:bench-stats}

\topic{Statistics}
\Cref{tab:bench-overview} summarizes \bench's 17 datasets
with a representative query each (full queries in
\Cref{tab:bench-overview-complete}).
All 104 queries span multiple databases; the maximum is
six databases across three systems (\ds{crmarenapro}),
and \ds{stockmarket} navigates 2{,}754 tables.
73\% of queries require reconciling ill-formatted join
keys. %
84\% require extracting structured values from
unstructured text. %
62\% require domain knowledge.
\crdy{\cdel{A detailed mapping between}\cnew{Appendix~\ref{append-sec:workload-prop} maps} each dataset and query to the four properties (\Cref{sec:bench-src-char}) and the \cdel{transformation techniques}\cnew{transformations} applied (\Cref{sec:bench-constr-method})\cdel{ are in Appendix~\ref{append-sec:workload-prop}}.}

\topic{Walk-through}
We illustrate \bench's structure using the \ds{bookreview}
dataset, which spans two databases (PostgreSQL and SQLite).
Consider the query {\em ``Which English-language books in
the `Literature \& Fiction' category have a perfect average
rating of 5.0?''}
To answer it, the agent receives a natural-language
description of each database and its schema:

\begin{tcolorbox}[
  colback=gray!5,
  colframe=gray!60,
  title={\scriptsize\textbf{Database Description (Excerpt)}},
  fonttitle=\scriptsize,
  boxrule=0.5pt,
  left=4pt, right=4pt, top=2pt, bottom=2pt,
  fontupper=\scriptsize
]
You are working with two databases to solve this query. Here are the descriptions of these two databases: \\[4pt]
\textbf{1. books\_database} \\
\textit{System:} 
PostgreSQL. 
Contains Amazon book information including descriptions, price, details, title, etc., up to 2023. \\
\textit{Tables:}
\begin{itemize}[nosep,leftmargin=12pt]
    \item \tinytt{books\_info} (Book information): \tinytt{title}, \tinytt{subtitle}, \tinytt{author}, \tinytt{rating\_number}, \tinytt{features}, \tinytt{description}, \tinytt{price}, \tinytt{store}, \tinytt{categories}, \tinytt{details}, \tinytt{book\_id}
\end{itemize}
\vspace{2pt}
\textbf{2. review\_database} \\
\textit{System:}
SQLite.
Contains Amazon book review information including ratings, review text, helpfulness votes, etc., up to 2023. \\
\textit{Tables:}
\begin{itemize}[nosep,leftmargin=12pt]
    \item \tinytt{review} (Review information): \tinytt{rating}, \tinytt{title}, \tinytt{text}, \tinytt{purchase\_id}, \tinytt{review\_time}, \tinytt{helpful\_vote}, \tinytt{verified\_purchase}
\end{itemize}
\vspace{2pt}
{\em (Descriptions for the table fields are omitted for brevity.)}
\end{tcolorbox}

It also receives a hints file describing the
transformations applied during dataset creation, which
flags data quality issues the agent may encounter:

\begin{tcolorbox}[
  colback=gray!5,
  colframe=gray!60,
  title={\scriptsize\textbf{Hints (Excerpt)}},
  fonttitle=\scriptsize,
  boxrule=0.5pt,
  left=4pt, right=4pt, top=2pt, bottom=2pt,
  fontupper=\scriptsize
]
\begin{itemize}[nosep,leftmargin=12pt]
    \item \tinytt{book\_id} (in \tinytt{books\_info}) and \tinytt{purchase\_id} (in \tinytt{review}) refer to the same book entities and can be matched via fuzzy join despite differences in formats.
    \item Some queries may require information from \tinytt{details} or \tinytt{categories} in \tinytt{books\_info}.
\end{itemize}
\end{tcolorbox}

\section{Experiments}
\label{sec:exp}

We evaluate three frontier LLM agents on \bench (5 trials
per query each); the best agent (GPT-5.5) reaches only
57\% pass@1, no agent exceeds 70\% pass@5, and one dataset
(\ds{patents}) is never solved. The rest of this section
covers the setup (\Cref{sec:exp-setup}), main results
(\Cref{sec:exp-result}), ablations on dataset descriptions
and hints (\Cref{sec:ablation}), and failure modes
(\Cref{sec:failure-analysis}). All agent trajectories are
released\del{ at }\footnote{\techreport{\url{https://drive.google.com/file/d/1SjCkvwsc4m1S17l_rzu9PHAAei3jAL4i/view?usp=drive_link}}\papertext{{\color{red!70!black}[URL redacted for submission]}}}.

\subsection{Setup}
\label{sec:exp-setup}

\topic{LLM Agents}
We test three frontier LLMs, each paired with a
state-of-the-art {\em harness}---an orchestration layer
that wraps an LLM in a loop with some tools and default
prompts, shown to substantially improve agent accuracy
over fixed LLM API calls~\cite{anthropic2025sweharness}.
We enable all built-in tools for each harness.
{\bf Claude-Sonnet-4.6} (Anthropic) uses the Claude Code
harness~\cite{anthropic2025claudecode,anthropic2025claudecodesdk},
and {\bf GPT-5.5} (OpenAI Azure Foundry) uses the Codex
harness~\cite{openai2025codex}. {\bf Gemini-3.1-Pro}
(Google)'s default harness, Gemini CLI, has no extensible
SDK~\cite{geminicli2025nosdk}, so we implement the
state-of-the-art ReAct harness~\cite{yao2022react} with
four tools---\ttt{list\_db}, \ttt{query\_db},
\ttt{execute\_python}, and \ttt{return\_answer}; full tool
signatures and prompt templates are in
\Cref{append-sec:prompt-templates}.
Individual tool calls time out after 600s; tool results
larger than 10{,}000 characters are written to disk so
agents can read them back.
All parameters are set to provider defaults. We disable
network access to prevent agents from loading external
data (e.g., \ds{agnews} ground-truth labels from Hugging
Face) to obtain correct answers.
\crdy{For brevity, \cdel{in the remainder of the paper, we omit the harness and use only the model name to denote the corresponding agent (i.e., model + harness)}\cnew{we hereafter name each agent by its model alone, omitting its harness}.}

\topic{Metrics}
We measure accuracy via pass@$k$~\cite{chen2021codex},
with $n=5$ trials per query per agent and $c$ correct:
$\text{pass@}k = 1 - \binom{n-c}{k}/\binom{n}{k}$.
All averages are {\em stratified}---per query, then per
dataset, then across the 17 datasets---so larger datasets
don't dominate. We also report cost (USD) and {\em
trajectory} statistics (iterations, tool calls). The
evaluation totals 4{,}680 trials at \$4{,}346.

\subsection{Results}
\label{sec:exp-result}

Our main experiments give each agent the full task
context: the natural-language question, the database
description, and the hints
(\Cref{sec:bench-constr-method}). We also run ablations
that remove the description and hints individually, presented in \Cref{sec:ablation}.

\begin{figure}[t]
    \centering
    \begin{minipage}{0.43\linewidth}
        \centering
        \raisebox{2pt}{\includegraphics[width=\linewidth]{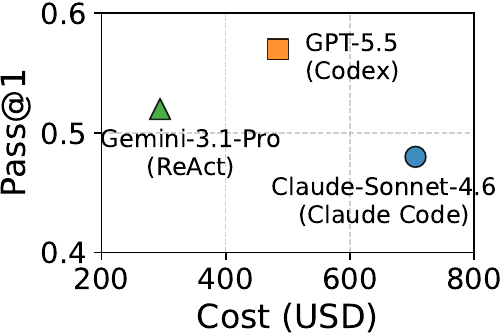}}
    \end{minipage}%
    \hfill
    \begin{minipage}{0.5\linewidth}
        \centering
        \includegraphics[width=\linewidth]{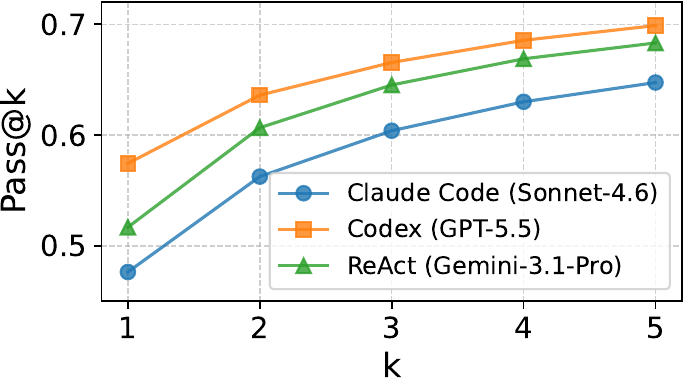}
    \end{minipage}

    \captionsetup{font=small}
    \begin{minipage}{0.43\linewidth}
        \captionof{figure}{Total cost (USD) vs.\ pass@1 per agent.}
        \label{fig:cost-acc}
    \end{minipage}%
    \hfill
    \begin{minipage}{0.53\linewidth}
        \captionof{figure}{Pass@$k$ as a function of $k$.}
        \label{fig:passk}
    \end{minipage}
    \vspace{-20pt}
\end{figure}

\topic{Accuracy}
\label{sec:exp-result-acc}%
\Cref{fig:cost-acc} plots pass@1 for each agent. GPT-5.5
leads at 57\%, followed by Gemini-3.1-Pro (52\%) and
Claude-Sonnet-4.6 (48\%)---Claude is the least accurate
despite being the most expensive.
\Cref{tab:acc_cost_per_dataset} reports per-dataset
breakdowns; no agent solves any query in \ds{patents}.
\Cref{fig:passk} plots pass@$k$ as a function of $k$:
agent rankings are stable, but the best agent still
{\bf \em reaches only 70\% pass@5}.

\begin{table}[t]
    \centering
    \caption{Per-agent pass@1 (Acc) and cost by dataset, in the main condition (description+hints). \best{Green}: highest pass@1; \zv{}: zero pass@1.
    }
    \label{tab:acc_cost_per_dataset}
    \scriptsize
    \setlength{\tabcolsep}{3pt}
    \renewcommand{\arraystretch}{0.92}
    \begin{tabular}{@{}l rr rr rr@{}}
    \toprule
    & \multicolumn{2}{c}{\bf Claude-Sonnet-4.6} & \multicolumn{2}{c}{\bf GPT-5.5} & \multicolumn{2}{c}{\bf Gemini-3.1-Pro} \\
    \cmidrule(lr){2-3}\cmidrule(lr){4-5}\cmidrule(lr){6-7}
    {\bf Dataset} & {\bf Acc} & {\bf Cost} & {\bf Acc} & {\bf Cost} & {\bf Acc} & {\bf Cost} \\
    \midrule
    \ds{agnews}             & \best{0.45} & \$19.9  & 0.40        & \$16.0  & 0.25        & \$8.0  \\
    \ds{bookreview}         & \best{0.93} & \$3.0   & \best{0.93} & \$3.1   & \best{0.93} & \$1.8  \\
    \ds{civic\_unstructured}& 0.32        & \$123.8 & 0.20        & \$68.3  & \best{0.34} & \$30.2 \\
    \ds{crmarenapro}        & 0.69        & \$25.9  & 0.68        & \$24.0  & \best{0.75} & \$17.5 \\
    \ds{cve}                & 0.08        & \$269.3 & \best{0.18} & \$149.5 & 0.12        & \$95.2 \\
    \ds{deps\_dev\_v1}      & \zv         & \$3.2   & \best{0.50} & \$6.1   & \best{0.50} & \$2.9  \\
    \ds{github\_repos}      & \best{0.45} & \$5.9   & 0.25        & \$10.3  & 0.40        & \$5.6  \\
    \ds{googlelocal}        & 0.50        & \$4.3   & \best{1.00} & \$4.0   & 0.80        & \$2.0  \\
    \ds{imdb}               & \best{0.76} & \$28.2  & 0.62        & \$49.5  & 0.38        & \$12.5 \\
    \ds{krama}              & 0.64        & \$105.8 & \best{0.74} & \$60.6  & 0.28        & \$47.8 \\
    \ds{music\_brainz\_20k} & 0.13        & \$2.3   & 0.67        & \$6.8   & \best{0.73} & \$2.4  \\
    \ds{pancancer\_atlas}   & \best{0.60} & \$5.1   & \best{0.60} & \$6.0   & 0.53        & \$1.8  \\
    \ds{patents}            & \zv         & \$7.2   & \zv         & \$14.2  & \zv         & \$4.5  \\
    \ds{stockindex}         & 0.80        & \$2.2   & \best{1.00} & \$3.2   & 0.93        & \$1.7  \\
    \ds{stockmarket}        & 0.80        & \$3.3   & \best{0.84} & \$3.6   & 0.80        & \$13.1 \\
    \ds{usaspending}        & 0.34        & \$84.7  & \best{0.50} & \$46.9  & 0.40        & \$39.6 \\
    \ds{yelp}               & 0.60        & \$11.3  & \best{0.66} & \$11.9  & 0.63        & \$7.6  \\
    \midrule
    {\bf Avg Acc; Total Cost}   & 0.48        & \$705.4 & \best{0.57} & \$484.0 & 0.52        & \$294.3 \\
    \bottomrule
    \end{tabular}
    \vspace{-10pt}
\end{table}

\topic{Cost-efficiency}
\label{sec:exp-result-cost}%
\Cref{fig:cost-acc} plots total API cost against pass@1
(per-dataset breakdown in \Cref{tab:acc_cost_per_dataset}).
Gemini-3.1-Pro offers the best
cost--accuracy tradeoff at \$294 total (52\% pass@1).
GPT-5.5 costs $1.6\times$ more for $+5$~pp accuracy, and
Claude-Sonnet-4.6 costs $2.4\times$ more yet is the
least accurate. \ds{cve} is the costliest dataset for every agent: it
contains 95~MB of unstructured vulnerability records that
agents pull into context to clean and join, and only
GPT-5.5 and Gemini-3.1-Pro answer a few of its queries
correctly.

\topic{Trajectory patterns}
\label{sec:exp-result-traj}%
\Cref{tab:latency-complexity} reports iterations
and tool calls per trajectory, broken down into database
queries
(\ttt{list\_db}/\ttt{query\_db} for ReAct; DB CLIs such as
\ttt{duckdb}, \ttt{sqlite3}, \ttt{psql}, \ttt{mongosh}, or
Python scripts importing only DB-related packages
(\ttt{duckdb}, \ttt{sqlite3}, \ttt{psycopg2},
\ttt{pymongo}) for Claude Code and Codex)
and Python executions
(\ttt{execute\_python} for ReAct; all other Python scripts
for Claude Code and Codex).
We highlight two takeaways.

\begin{table}[t]
    \centering
    \caption{Trajectory statistics per agent.
    Each metric is first averaged across 5 trials for a given query,
    then across queries. The hardest queries for the agents all come from \ds{cve}.}
    \label{tab:latency-complexity}
    \scriptsize
    \setlength{\tabcolsep}{3pt}
    \renewcommand{\arraystretch}{0.92}
    \begin{tabular}{
    >{\raggedright\arraybackslash}p{0.40\columnwidth}
    >{\raggedleft\arraybackslash}p{0.14\columnwidth}
    >{\raggedleft\arraybackslash}p{0.12\columnwidth}
    >{\raggedleft\arraybackslash}p{0.10\columnwidth}
    >{\raggedleft\arraybackslash}p{0.12\columnwidth}
    }
    \toprule
    {\bf Agent} & {\bf \#LLM calls} & {\bf \#Tool calls} & {\bf \#DB queries} & {\bf \#Python execs} \\
    \midrule
    \multicolumn{5}{l}{\cellcolor{gray!10}\em Stratified average across all queries} \\
    Claude-Sonnet-4.6 & 30.5 & 30.3 & 10.3 & 11.2\\
    GPT-5.5 & 16.3 & 12.7 & 9.2 & 2.3\\
    Gemini-3.1-Pro & 23.4 & 24.9 & 10.1 & 13.8\\
    \multicolumn{5}{l}{\cellcolor{gray!10}\em Hardest query (highest per-query average)} \\
    Claude-Sonnet-4.6 & 160.0 & 159.0 & 38.5 & 102.5\\ %
    GPT-5.5 & 44.4 & 40.0 & 14.0 & 21.4\\ %
    Gemini-3.1-Pro & 73.2 & 74.6 & 33.4 & 40.2\\ %
    \bottomrule
    \end{tabular}
    \vspace{-10pt}
\end{table}

\takeaway{Writing larger programs upfront is more cost-efficient than step-by-step exploration}
GPT-5.5 typically solves each query with one large Python
program that inspects schemas, queries the database, and
aggregates results in a single process. Claude-Sonnet-4.6
splits the same work into many smaller commands: a 0.9:1
DB-to-Python ratio versus 4:1 for GPT-5.5. For instance,
on \ds{cve}/query1, Claude-Sonnet-4.6 needs five Python
executions to explore the database while GPT-5.5 does so
in one. This pattern largely explains the cost gap:
Claude-Sonnet-4.6 costs \$705 at 48\% pass@1, while
GPT-5.5 costs \$484 at 57\%.
Note that Gemini-3.1-Pro's DB-to-Python ratio is not
directly comparable, since its ReAct harness exposes
Python as a dedicated tool whereas Claude Code and Codex
invoke Python via bash.

\takeaway{Parallel tool calling is underutilized, but could improve latency and cost}
Multi-database workloads in \bench would benefit from
issuing tool calls in parallel, since each source can be
queried independently. Exploratory calls (schema lookups,
sample queries) are an especially natural fit for
parallelization, and they make up a large fraction of
trajectories: 51\% for GPT-5.5, 39\% for Claude-Sonnet-4.6,
and 22\% for Gemini-3.1-Pro (\Cref{tab:exploration}).
A single iteration may issue multiple tool calls in
parallel, which is why some agents in
\Cref{tab:latency-complexity} have fewer iterations than
tool calls. However, all three agents parallelize on fewer
than 10\% of turns: Claude-Sonnet-4.6 most often (8.2\%),
followed by GPT-5.5 (5.2\%) and Gemini-3.1-Pro (5.1\%).
\techreport{When they do parallelize, GPT-5.5 and Gemini-3.1-Pro
occasionally issue large bursts---up to 16 and 100
concurrent calls in a single turn.}

\crdy{We also ran a \cdel{shared harness}\cnew{shared-harness} experiment to evaluate different models\cdel{. We use}\cnew{, pairing} ReAct with GPT 5.2, Claude Opus 4.6, and Gemini 3.1 Pro. Due to budget limits, we \cdel{evaluated about half of the queries; 54 queries across 12 datasets (selecting the least expensive queries)}\cnew{used the 54 least expensive queries across 12 datasets}, with five trials per model per query. \cdel{GPT 5.2 achieves 29\% pass@1 and 44\% pass@5, Claude Opus 4.6 achieves 55\% and 69\%, and Gemini 3.1 Pro achieves 60\% and 75\%. Detailed results are in Appendix~\ref{append-sec:shared-harness-results}.}\cnew{Pass@1/pass@5 is 29\%/44\% for GPT 5.2, 55\%/69\% for Claude Opus 4.6, and 60\%/75\% for Gemini 3.1 Pro (details in Appendix~\ref{append-sec:shared-harness-results}).}}

\subsection{Ablations}
\label{sec:ablation}

The main experiments give each agent both dataset
descriptions (databases, tables, schemas) and ``hints''
that describe our induced data quality issues. Here, we perform two
ablations: removing hints only (``w/o hints'') and
removing both descriptions and hints (``w/o
desc+hints''), with 5 trials per query in each.

\begin{table}[t]
  \centering
  \caption{Average pass@1, total cost (USD), and per-trial tool calls under each ablation, with the difference vs.\ the full-info setting in parentheses. \best{Green}: highest pass@1 across agents per row.}
  \label{tab:abl_acc_cost_summary}
  \scriptsize
  \setlength{\tabcolsep}{2pt}
  \renewcommand{\arraystretch}{0.95}
  \begin{tabular*}{\columnwidth}{@{\extracolsep{\fill}}l rrr@{}}
    \toprule
    {\bf Setting} & {\bf Acc} & {\bf Cost} & {\bf \#Tools} \\
    \midrule
    \multicolumn{4}{@{}l}{\em Claude-Sonnet-4.6} \\
    w/o hints       & 0.50 (+0.02)        & \$670 (-\$35) & 29.0 (-1.3) \\
    w/o desc+hints  & 0.43 (-0.05)        & \$686 (-\$19) & 37.9 (+7.6) \\
    \midrule
    \multicolumn{4}{@{}l}{\em GPT-5.5} \\
    w/o hints       & \best{0.58} (+0.01) & \$483 (-\$1)  & 11.3 (-1.4) \\
    w/o desc+hints  & \best{0.55} (-0.02) & \$453 (-\$31) & 15.8 (+3.1) \\
    \midrule
    \multicolumn{4}{@{}l}{\em Gemini-3.1-Pro} \\
    w/o hints       & 0.48 (-0.04)        & \$280 (-\$14) & 25.6 (+0.7) \\
    w/o desc+hints  & 0.47 (-0.05)        & \$299 (+\$5)  & 34.2 (+9.3) \\
    \bottomrule
  \end{tabular*}
  \vspace{-10pt}
\end{table}

\topic{Accuracy and cost}
\Cref{tab:abl_acc_cost_summary} reports the average pass@1, total cost, and per-trial tool calls under each ablation; the per-dataset breakdown is in \Cref{append-sec:abl_results}.
Removing hints or descriptions sometimes decreases accuracy (as expected, since the agent has less information) and sometimes increases it: hints apply to the entire dataset, so when a particular query doesn't need a particular hint, the agent can be ``misled''. For example, on \ds{crmarenapro}/query12, the agent sees a hint about join-key corruption---which actually applies to other queries in this dataset, not query12---and assumes its initial (correct) answer is wrong because it didn't clean for that corruption, so it reports a different, incorrect answer. Under ``w/o hints'', the agent has no such suspicion and answers correctly in 2/5 trials. Across the board, however, accuracy shifts are marginal: pass@1 moves by at most $5$~pp in either direction. Costs also shift only modestly, between $-6\%$ and $+2\%$ across agents and conditions.

\topic{Tool use}
Unsurprisingly, removing the description (``w/o desc+hints'') drives all three agents to issue more tool calls on average---$+3$ to $+9$ per trial---as they probe schemas and surface data quality issues themselves. Removing hints alone has little effect ($\pm 1$ call).
Interestingly, sometimes removing hints actually {\em decreases} exploration. For example, on \ds{usaspending}, where a join column is corrupted in ways that require row-by-row reasoning, in the main experiment, agents consider the hint and issue many tool calls to do that reasoning (Claude-Sonnet-4.6: 53.5/trial, Gemini-3.1-Pro: 48.9/trial). But under ``w/o hints'', they never discover the corruptions, issue far fewer tool calls (39.4 and 22.7 respectively), and return the wrong answer.

\subsection{Error Analysis}
\label{sec:failure-analysis}

We also analyze failed trajectories to identify where the
agents are breaking down.

\topic{Methodology}
\label{sec:fm-method}%
Two authors examined 30 failed trajectories from the
\ds{bookreview} dataset and defined five failure modes
specific to data agents. To classify
presence of failure modes at scale, we used GPT-5 as an
LLM judge\del{---following prior
work~\cite{cemri2025multi,merrill2026terminal}---}over one
incorrect trajectory per query per agent in the main
experiment, yielding 196 annotated trajectories. 
\crdy{The LLM judge doesn't perform open ended error analysis; it determines whether a trajectory exhibits each human defined failure mode, following prior work in agent evaluation~\cite{cemri2025multi, merrill2026terminal, zhuge2024agent}. We follow the procedure used by Terminal Bench.}
The full
annotation prompt and per-agent breakdown are in
\Cref{append-sec:fm-annot-prompt}, and trajectories
illustrating each FM appear in
\Cref{append-sec:fm-examples}. The failure modes are:

\label{sec:fm-taxonomy}%
\begin{itemize}[leftmargin=0pt, itemindent=*, labelsep=4pt, align=left, nosep]
\item \fmbadge{FM1} \emph{Fails before planning}: the agent makes no attempt---either returning \ttt{None} as its tool call (\fmbadge{FM1\,(no\_tool\_call)}) or refusing without invoking any tool (\fmbadge{FM1\,(other)}).
\item \fmbadge{FM2} \emph{Incorrect plan}: the agent's plan cannot produce the correct answer even if executed perfectly---e.g., averaging book-level averages within a decade rather than averaging the underlying review ratings directly.
\item \fmbadge{FM3} \emph{Incorrect data selection}: the plan is right but the agent queries the wrong source---e.g., checking the \ttt{description} column for a book's language when the \ttt{details} column holds it.
\item \fmbadge{FM4} \emph{Incorrect implementation}: plan and data are right but execution is wrong---e.g., a regex intended to extract a publication year inadvertently matches an ISBN segment from \ttt{details}.
\item \fmbadge{FM5} \emph{Runtime error}: API failures (non-200 HTTP responses, tool-call array overflows, ``service unavailable'') or request timeouts ($>$600s).
\end{itemize}

\topic{Results}
\label{sec:fm-results}%
\Cref{fig:fm-freq} and \Cref{fig:fm234-freq} in \Cref{append-sec:fm-freq-plots} show the failure-mode breakdown across all 1{,}560 trajectories in the main experiment. We report two takeaways below.

\takeaway{Agents typically select the right data, but fail at planning the computation or implementing it correctly}
Among the trajectories that completed but returned
incorrect answers (\Cref{fig:fm234-freq}),
\fmbadge{FM4} (incorrect implementation) is the most
common at 57\%, followed by \fmbadge{FM2}
(incorrect plan) at 31\%, \fmbadge{FM3}
(incorrect data selection) at 11\%,
and \fmbadge{FM1 (other)} at 1\%.
The low rate of \fmbadge{FM3} indicates that
agents generally identify the correct tables and columns;
the dominant challenge is deciding what to compute
and computing it correctly.
Together, \fmbadge{FM2} and \fmbadge{FM4} account
for 90\% of incorrect answers.

\takeaway{All agents rely on regex for text extraction and fail when it isn't expressive enough}
When agents need to parse text fields---e.g., pulling a
date out of a sentence---they default to authoring and
executing a regex; only Claude-Sonnet-4.6 ever calls an
LLM-based extractor (and only on \ds{agnews}). This
breaks down on \bench's text-heavy datasets. For example,
on \ds{patents} (0\% pass@1 for every agent), dates appear
in many natural-language forms (e.g., ``dated 5th March
2019'', ``March the 18th, 2019''), and no regex the agents
write covers all of them. In another example, on
\ds{pancancer\_atlas}, a regex for \ttt{MALE} matches
inside \ttt{FEMALE} and mislabels gender. Perhaps exposing
dedicated extraction tools (e.g., date parsers, NER,
LLM-based extractors) as first-class tools, rather
than forcing agents to author them ad hoc, could unlock
many of \bench's hardest queries.

\crdy{While our quantitative analysis is intended to compare the {\em relative} prevalence of the failure modes---rather than estimate the {\em exact} fraction of trajectories exhibiting each failure mode---to further adjudicate GPT-5's judge and validate the \cdel{aforementioned }takeaways\cnew{ above}, we \cdel{did two additional experiments as follow}\cnew{ran two more experiments}:
\begin{itemize}[leftmargin=0pt, itemindent=*, labelsep=4pt, align=left, nosep]
\item Human validation: One author \cdel{manually }reviewed 30 trajectories, 10 \cdel{sampled from each}\cnew{per} agent. 20 labels\cdel{, or 67\%,}\cnew{ (67\%)} match the GPT-5 judge exactly; 3\cdel{, or 10\%,}\cnew{ (10\%)} partially match because the human assigns \cdel{additional}\cnew{extra} failure modes; and 7\cdel{, or 23\%,}\cnew{ (23\%)} disagree. Importantly, the human labels \cdel{produce the same ranking of failure mode prevalence as the LLM}\cnew{rank failure-mode prevalence the same as the} judge.
\item Agreement across LLM judges: We \cdel{independently annotated}\cnew{re-annotated} all trajectories \cdel{using}\cnew{with} GPT-5.4 \cdel{(in addition to GPT-5 in the paper), }to measure judge agreement. Cohen’s $\kappa$ is 0.80 for FM1 and \cdel{0.42 to 0.46}\cnew{0.42--0.46} for \cdel{FM2, FM3, and FM4}\cnew{FM2--FM4}, indicating strong or moderate agreement. More importantly\cdel{ for our analysis}, both judges \cdel{produce}\cnew{give} the same ordering: FM2\cdel{, 42\% with GPT-5.4 and 31\% with GPT-5, and FM4, 44\% and 57\%, dominate; FM3 follows at 13\% and 11\%; and FM1 is rare at 1\% for both judges.}\cnew{ (42\% with GPT-5.4, 31\% with GPT-5) and FM4 (44\%, 57\%) dominate, FM3 follows (13\%, 11\%), and FM1 is rare (1\% for both).}
\end{itemize}
}

\section{Conclusion}
\label{sec:conclusion}
We presented \bench, the first benchmark for evaluating
data agents on realistic, multi-database queries.
Grounded in a formative study of enterprise workloads,
\bench comprises 104 queries across 17 datasets and
4 database systems.
Even the best frontier agent achieves only 57\% pass@1,
and 90\% of failures stem from incorrect plans or
incorrect implementations rather than data selection.
\bench represents a significant step forward for data
agent evaluation, and we hope it helps the community
build data agents that can be trusted in production.

\section{Limitations}
\label{sec:limitations}

\bench reflects several tradeoffs we made during
construction.
First, \bench uses four open-source database systems
(PostgreSQL, MongoDB, SQLite, and DuckDB) so that it can
be run without commercial licenses; enterprise deployments
commonly involve proprietary systems such as Oracle,
Snowflake, and Databricks, which introduce additional
challenges (e.g., vendor-specific SQL dialects,
access-control policies) that \bench does not yet capture.
Second, all queries are analytical and read-only---the
benchmark does not cover transactional workloads in which
agents must insert, update, or delete records.
\crdy{In fact, data systems typically use separate benchmarks for distinct workload classes rather than a single benchmark intended to cover them all. For example, analytical and transactional workloads are conventionally evaluated using TPC-H and TPC-C, respectively. Hex, a popular data agent startup, has multiple internal benchmarks for different workloads, such as warehouse analytics~\cite{hexwarehouse} and chart generation~\cite{hexchart}. 
}
Third, every query has a single deterministic answer,
which enables fully automatic evaluation but excludes
open-ended analytical questions
(\Cref{sec:bench-src-char}) and queries requiring
statistical modeling or machine-learning workflows;
existing data science benchmarks~\cite{lai2023ds,
huang2024code, lai2025kramabench} already target these
capabilities.
Fourth, \bench excludes queries that depend on external
APIs (email clients, web search, third-party data
providers); our formative study found these common in
production, but live endpoints break reproducibility.
\crdy{Moreover, live external data introduce data quality issues that we cannot continuously identify and correct, making it difficult to evaluate agents fairly or ensure that the data meets the expectations established in our formative studies.}
Fifth, \bench contains 104 queries across 17
datasets---comparable in size to other carefully curated
benchmarks (e.g., FrontierMath~\cite{glazer2024frontiermath},
TerminalBench~\cite{merrill2026terminal})---but does not
yet cover every domain or database system encountered in
production.
Despite manual verification and independent review by
different authors (\Cref{sec:task-formulation}), some
queries may be ambiguous or hints may contain
inconsistencies. We plan to expand \bench over time and
welcome community contributions and corrections.
\crdy{We hope the community develops benchmarks for additional data agent workloads, and we plan to extend DAB to cover more workloads over time.}

\section{Ethical Considerations}

Our formative-study interviews and data-handling protocol
were approved by our institution's IRB; participants gave
informed consent, and we release no proprietary customer
data, schemas, or queries. \bench is built from
open-source datasets used consistent with their licenses
(\Cref{sec:bench-constr-method}) and contains no PII. We
release \bench's transformation scripts, validation
scripts, and evaluation code under the MIT license; the
transformed datasets are released for research use only,
consistent with the licenses of their underlying sources.

We used LLMs at several points in this work---GPT-4o to
embed structured attributes into free-form text during
dataset construction (with all transformations
author-verified to preserve deterministic ground truth),
GPT-5 to annotate which failure modes appear in each
incorrect trajectory (\Cref{fig:fm234-freq}), and Claude
Haiku to classify tool calls as exploratory or analytical
(\Cref{tab:exploration}).

The total experiment cost is \$4{,}346
(\Cref{sec:exp-result-cost}). This is a substantial figure,
especially for university research labs, and it is a real
barrier to reproducibility: re-running the full
evaluation, or evaluating a new agent on \bench, takes
thousands of dollars. To partially mitigate this, we
release all agent trajectories and code so that downstream
analyses can be performed without re-running the
experiments.

We caution that accuracy on \bench does not imply
deployment readiness, particularly in regulated domains.

\section*{Acknowledgments}
We acknowledge support from grants DGE-2243822, IIS-2129008, IIS-1940759, and IIS-1940757 awarded by the National Science Foundation, funds from the State of California, funds from the Alfred P. Sloan Foundation, as well as EPIC lab sponsors: Adobe, Google, G-Research, Microsoft, PromptQL, Sigma Computing, Snowflake, Bridgewater, and Jane Street. Compute credits were provided by Azure, Modal, NSF (via NAIRR), OpenAI.

\bibliography{sample}

\clearpage

\appendix

\section{Benchmark Construction Details}
\label{append-sec:bench-details}

\subsection{Interview Protocol and Sample Response}
\label{append-sec:interview-protocol}

We solicited each enterprise's data-agent workload through
a structured questionnaire. For each enterprise, we asked
the analysts and engineers building the data agent to
provide:
\begin{enumerate}[leftmargin=*, nosep]
    \item the natural-language queries their end users
    pose (or wish to pose) to a data agent;
    \item the schemas of their data sources, including the
    database system each source uses (e.g., PostgreSQL,
    MongoDB, Salesforce, vector store);
    \item where each piece of data is stored, including
    overlapping or redundant data across systems;
    \item the join keys between tables, both within a
    single database and across databases; and
    \item for each example query, a plan describing how an
    analyst would manually perform the joins,
    transformations, and any text extractions needed to
    answer it.
\end{enumerate}

Below we reproduce an anonymized sample response from one
enterprise, an enterprise project-management SaaS platform
(comparable in scope to products such as Asana or Monday).

\paragraph{Data sources.}
\begin{enumerate}[leftmargin=*, nosep]
    \item \textbf{Platform database} (PostgreSQL):
    \ttt{customers}, \ttt{users}, \ttt{projects},
    \ttt{tasks}, \ttt{feature\_usage}.
    \item \textbf{Support database} (PostgreSQL):
    \ttt{agents}, \ttt{tickets}, \ttt{ticket\_responses}.
    \item \textbf{Salesforce}: \ttt{accounts},
    \ttt{opportunities}, \ttt{contacts}.
    \item \textbf{Billing database} (MongoDB):
    \ttt{invoices}, \ttt{usage\_metrics}.
    \item \textbf{Documentation} (vector database):
    \ttt{documents} with text and embeddings.
\end{enumerate}
The response also provided example rows and join keys for
each source (e.g., \ttt{customers.customer\_id}~$\leftrightarrow$~\ttt{users.customer\_id};
Salesforce \ttt{accounts.Name} fuzzily matches
\ttt{customers.company\_name}).

\paragraph{Example queries.}
The response provided dozens of queries organized into
five categories. We list a representative example from
each, with the analyst's stated plan.

\paragraph{(1) Joining across data sources.}

\emph{Exact (1:1) join.} ``Which enterprise-plan customers
have raised a support ticket in the last 30 days?''
\textit{Plan:} filter \ttt{customers} by
\ttt{plan\_type=enterprise}, join with \ttt{users} on
\ttt{customer\_id}, query \ttt{tickets} by these
\ttt{user\_id}s with \ttt{created\_date} in the last 30
days.

\emph{Programmatic transform join.} ``Which Salesforce
accounts correspond to our platform customers?''
\textit{Plan:} normalize company names in both
\ttt{customers} and Salesforce \ttt{accounts} (strip
``Inc'', ``Corp'' suffixes; lowercase); join on normalized
names.

\emph{Fuzzy/semantic join.} ``Which customers mentioned in
Salesforce opportunities are the same as our existing
platform customers?'' \textit{Plan:} fuzzy string matching
to reconcile variations such as ``International Business
Machines'' vs.\ ``IBM''.

\paragraph{(2) Bulk semantic operations.}
\emph{Classification.} ``Does our average ticket
resolution time differ for production vs.\
non-production issues?''
\textit{Plan:} classify each ticket from
\ttt{tickets.description} using an LLM; aggregate
\ttt{resolution\_time} per class.

\emph{Extraction.} ``Which software versions have
reported the most issues?''
\textit{Plan:} extract version numbers from ticket
descriptions (e.g., ``v2.1.3'') with regex/NER; group by
version.

\emph{Clustering.} ``What are the top issues that keep
recurring?''
\textit{Plan:} semantically cluster ticket descriptions
and rank clusters by frequency.

\emph{Generation/summarization.} ``Generate a monthly
support performance report with key insights.''

\paragraph{(3) Search.}
``If someone is running into XYZ error, what should we
do?''
\textit{Plan:} vector-search the documentation database and
the resolved-ticket archive; combine results.

\paragraph{(4) Domain knowledge.}
``Which deals are `at risk'?''
\textit{Plan:} apply this company's definition of ``at
risk''---opportunities with \ttt{Probability < 50\%}
closing within 30 days, or accounts with billing issues or
declining usage metrics.

\paragraph{(5) Open-ended.}
``What should I do to improve my support process?''
(Excluded from \bench because no single correct answer.)

\subsection{All Queries in \bench}
\label{append-sec:all-queries}
\autoref{tab:bench-overview-complete} lists all queries in
\bench. The raw data are available at \techreport{\url{https://github.com/ucbepic/DataAgentBench}}\papertext{{\color{red!70!black}[URL redacted for submission]}}.

\clearpage
\onecolumn
\begin{center}
    \scriptsize
    \setlength{\tabcolsep}{5pt}
    \renewcommand{\arraystretch}{1.3}
    \begin{longtable}{
    @{}
    p{0.10\columnwidth}
    >{\itshape}p{0.84\columnwidth}
    @{}
    }
    \caption{All queries in \bench} \label{tab:bench-overview-complete}\\
        \toprule
        {\bf Dataset} & {\bf Query}\\
        \midrule
        \endfirsthead
        \multicolumn{2}{c}%
        {\bf {\tablename\ \thetable{} (continued)}} \\
        \toprule
        {\bf Dataset} & {\bf Query}\\ 
        \midrule
        \endhead
        \bottomrule
        \endfoot
        \bottomrule
        \endlastfoot
        
        \multirow[t]{4}{*}{\dst{agnews}} & {(1) What is the title of the sports article whose description has the greatest number of characters?} \\*
        \cmidrule{2-2}
         & {(2) What fraction of all articles authored by Amy Jones belong to the Science/Technology category?} \\*
        \cmidrule{2-2}
         & {(3) What is the average number of business articles published per year in Europe from 2010 to 2020, inclusive?} \\*
        \cmidrule{2-2}
         & {(4) In 2015, which region published the largest number of articles in the World category?} \\
        \hline
        \multirow[t]{3}{*}{\dst{bookreview}} & {(1) Which decade of publication (e.g., 1980s) has the highest average rating among decades with at least 10 distinct books that have been rated? Return the decade with the highest average rating.} \\*
        \cmidrule{2-2}
         & {(2) Which English-language books in the `Literature \& Fiction' category have a perfect average rating of 5.0? Return all matching books.} \\*
        \cmidrule{2-2}
         & {(3) Which books categorized as `Children's Books' have received an average rating of at least 4.5 based on reviews from 2020 onwards?} \\
        \hline
        \multirow[t]{10}{*}{\dst{civic\_unstructured}} & {(1) As of 2023-01-01, list all projects whose most recent report (dated on or before 2023-01-01) records both a type of `capital' and a status of `design' and that have accumulated total funding of more than \$500,000. Accumulated total funding means the sum of all funding amounts granted to that project on or before 2023-01-01.} \\*
        \cmidrule{2-2}
         & {(2) As of 2023-01-01, among all projects whose most recent report (dated on or before 2023-01-01) records a status of `completed', which project has the highest accumulated total funding, and what is that funding amount in USD?} \\*
        \cmidrule{2-2}
         & {(3) As of 2022-08-01, list all funding sources where at least 25\% of the distinct projects they have funded on or before that date have a status of `completed' in their most recent report dated on or before 2022-08-01.} \\*
        \cmidrule{2-2}
         & {(4) What is the ratio of total funding amount granted to capital projects versus disaster projects between 2021-06-01 and 2022-06-01 (inclusive)? For each funding record, use the project's type as of the date that funding was granted. Express as a single decimal number rounded to 2 decimal places (capital\_total / disaster\_total).} \\*
        \cmidrule{2-2}
         & {(5) Between 2021-06-01 and 2023-01-01, for all projects that underwent at least one status change within this period, what is the average total funding received per project? Report the average in USD, rounded to 2 decimal places.} \\*
        \cmidrule{2-2}
         & {(6) Across all years in the funding records, which year did capital projects receive the highest total funding amount? Report the year and the total funding amount in USD.} \\*
        \cmidrule{2-2}
         & {(7) As of 2022-11-01, list all projects whose most recent report records both a type of `disaster' and a status of `not started' and that have accumulated total funding of more than \$1,000,000.} \\*
        \cmidrule{2-2}
         & {(8) As of 2022-06-01, among all projects whose most recent report records a type of `capital' and a status of `construction', which project has the lowest accumulated total funding, and what is that funding amount in USD?} \\*
        \cmidrule{2-2}
         & {(9) As of 2022-11-01, list all funding sources where more than 55\% of the distinct projects they have funded have a type of `disaster' in their most recent report dated on or before 2022-11-01.} \\*
        \cmidrule{2-2}
         & {(10) What is the ratio of total funding granted to projects with a status of `completed' versus projects with a status of `not started' between 2021-09-01 and 2022-09-01 (inclusive)? Express as a single decimal number rounded to 2 decimal places (completed\_total / not\_started\_total).} \\
        \hline
        \multirow[t]{13}{*}{\dst{crmarenapro}} & {(1) Can this lead be qualified based on the latest discussions? If the answer is no, which factors---`Budget', `Authority', `Need', or `Timeline'---are responsible? Return only one or several of the four BANT factors that the lead qualification fails to meet (i.e. `Budget', `Authority', `Need', `Timeline').\newline \newline \#\# Lead qualification guide.\newline Look for the voice call transcripts with the lead and relevant knowledge articles to justify the lead qualification.\newline \newline - Lead Id to be considered is: 00QWt0000089AekMAE} \\*
        \cmidrule{2-2}
         & {(2) Does the cost and setup of this quote comply with our company policy? If it doesn't, which knowledge article is it in conflict with? Return only the Id of the knowledge article that the quote violates. If no violation is found, return None.\newline \newline \#\# Quote approval guide.\newline Look for relevant knowledge articles to justify the quote approval.\newline \newline - Quote Id to be considered is: 0Q0Wt000001WSDVKA4} \\*
        \cmidrule{2-2}
         & {(3) Is the stage name accurately representing the tasks for this opportunity? If it is not, what should the appropriate stage name be? Return only the correct stage label among (`Qualification', `Discovery', `Quote', `Negotiation', `Closed/).\newline \newline - Opportunity Id to be considered is: 006Wt000007BGGjIAO} \\*
        \cmidrule{2-2}
         & {(4) Is there a particular month in the past 10 months where the number of SecureAnalytics Pro cases significantly exceeds those of other months? The associated product Id is 01tWt000006hVJdIAM. Return only the month name.\newline \newline - Today's date: 2021-04-10} \\*
        \cmidrule{2-2}
         & {(5) What has been the most frequent problem AI Cirku-Tech encountered over the past five months? The associated product Id is 01tWt000006hV8LIAU. Return only the issue Id of the most reported issue for this product.\newline \newline - Today's date: 2023-01-16} \\*
        \cmidrule{2-2}
         & {(6) Is the product setup in this quotation, including elements like quantity and price, against company regulations? Return only the Id of the knowledge article that the invalid config violates.\newline \newline \#\# Invalid config guide.\newline Look for the relevant knowledge articles to justify the invalid config.\newline \newline - Quote Id to be considered is: 0Q0Wt000001WRAzKAO} \\*
        \cmidrule{2-2}
         & {(7) Did the agent breach the policy, and if so, which knowledge article was breached? Return only the Id of the knowledge article or None if no violation is found.\newline \newline - Case Id to be considered is: 500Wt00000DDyznIAD} \\*
        \cmidrule{2-2}
         & {(8) Identify the agent with the fewest transfer counts in the last 4 quarters among those who handled more than 0 cases. Return only the Id of the agent.\newline \newline \#\# Transfer Count Policy\newline - Definition: The number of instances a case was reassigned or transferred from one agent to another. Each transfer from agent A to agent B adds to the transfer count for agent A.\newline - In the queries that specify `agents managed/queries x cases'---this filter applies to both the first agent that the case was first assigned to and the agent that the case was transferred to. This means that if an agent has 2 cases that was initially assigned to itself by admin and 1 case transferred from another agent, a filter like `handled/managed at least 3 cases'' would not filter this agent out.\newline - For cases that have NOT been transferred to an other agent, there will be only ONE `Owner Assignment', and for those that have been transferred, there will be MORE THAN ONE `Owner Assignment'.\newline \newline \#\# Today's date: 2023-04-10} \\*
        \cmidrule{2-2}
         & {(9) Which states have the quickest case closure time in the past 6 quarters? Return only the two-letter abbreviation of the most matching state (eg. CA).\newline \newline - Today's date: 2022-10-26} \\*
         \cmidrule{2-2}
         & {(10) In the past four months, which agent had the lowest average handle time for those processing more than one case? Return only the Id of the agent.\newline \newline \#\# Handle Time Policy\newline - Definition: The duration taken to close a case. Specifically, it is the time from when a case is opened to when it is closed.\newline - In the queries that specify `agents managed/queries x cases'---this filter applies to both the first agent that the case was first assigned to and the agent that the case was transferred to. This means that if an agent has 2 cases that was initially assigned to itself by admin and 1 case transferred from another agent, a filter like `handled/managed at least 3 cases' would not filter this agent out.\newline - When computing handle time, we do not compute handle time for cases that have been transferred to other agents.\newline - For cases that have NOT been transferred to an other agent, there will be only ONE `Owner Assignment', and for those that have been transferred, there will be MORE THAN ONE `Owner Assignment'.\newline \newline \#\# Today's date: 2023-09-02} \\*
        \cmidrule{2-2}
         & {(11) Can you show me the AI processing unit I purchased last month? Return only the Id of the product from the contact's relevant past transaction.\newline \newline - Contact Id interacting: 003Wt00000Jqy8SIAR\newline - Today's date: 2021-07-15} \\*
        \cmidrule{2-2}
         & {(12) Who had the quickest average turnaround from opening to closing opportunities among agents in April 2023? Return only the Id of the agent.\newline \newline \#\# Sales Cycle Policy\newline - Definition: The sales cycle is measured as the number of days between an opportunity's creation date and the company signed date on the corresponding contract.\newline \newline \#\# Today's date: 2024-09-12} \\*
        \cmidrule{2-2}
         & {(13) Identify the agent who achieved the top sales figures for orders made in the past five months. Return only the Id of the agent.\newline \newline \#\# Sales Amount Policy\newline - Definition: The sales amount for an order is calculated as the product of the quantity and the unit price from the Order object (Quantity * UnitPrice). An opportunity is eligible if its associated contract has a company signed date that falls within the time interval of interest.\newline \newline - Today's date: 2022-11-25} \\
        \hline
        \multirow[t]{10}{*}{\dst{cve}} & {(1) How many distinct CVEs published in 2023 are listed in CISA's Known Exploited Vulnerabilities (KEV) catalog AND have at least one vulnerable Apache product configuration?} \\*
        \cmidrule{2-2}
         & {(2) Among CVEs in CISA's KEV catalog flagged as having known ransomware-campaign use, identify the canonical vendor (using kev\_vendor\_aliases to resolve KEV vendor\_project surface forms) whose CVEs include the most CRITICAL-severity entries (consider only vendors with at least 3 KEV-ransomware-flagged CVEs).} \\*
        \cmidrule{2-2}
         & {(3) How many entries in CISA's KEV catalog reference a CVE identifier for which there is NO matching record in the NVD vulnerability database (vulns\_db)?} \\*
        \cmidrule{2-2}
         & {(4) For each canonical vendor that has at least 10 distinct CVEs both listed in CISA's KEV catalog AND with at least one vulnerable CPE configuration, compute the share of those CVEs whose CVSS v3 base score is 9.0 or higher. Return the vendor with the highest share, breaking ties alphabetically.} \\*
        \cmidrule{2-2}
         & {(5) Among CVEs in CISA's KEV catalog that affect Apache Tomcat (canonical vendor `apache', product `tomcat'), how many distinct affected version strings are recorded, excluding wildcard (`*') and null values? Treat versions differing only in encoding as the same version (e.g., `9.0.71', `9,0,71', and `v9\_0\_71').} \\*
        \cmidrule{2-2}
         & {(6) How many distinct CVEs have NO English-language description, DO have at least one non-English description, AND are listed in CISA's KEV catalog?} \\*
        \cmidrule{2-2}
         & {(7) Among CPE match entries flagged as vulnerable AND whose CVE is listed in CISA's KEV catalog, which canonical (vendor, product) pair has the highest count of rows? Return in the form `vendor:product' (both lowercase).} \\*
        \cmidrule{2-2}
         & {(8) How many distinct CVEs are listed in CISA's KEV catalog AND have at least one vulnerable Microsoft product configuration AND have a CVSS v3 base score of 7.0 or higher?} \\*
        \cmidrule{2-2}
         & {(9) How many distinct CVEs appear in more than one row of the cves table in vulns\_db with conflicting cvss3\_attack\_vector values AND are listed in CISA's KEV catalog?} \\*
        \cmidrule{2-2}
         & {(10) Among the products listed in CISA's KEV catalog whose canonical vendor is Microsoft, how many distinct products have at least one associated CVE with a CVSS v3 base score of 9.0 or higher? Split rows whose product field contains comma-separated lists into individual products, and count case-insensitively.} \\
        \hline
        \multirow[t]{2}{*}{\dst{DEPS\_DEV\_V1}} & {(1) Considering only the latest release versions for each distinct NPM package, which packages are the top 5 most popular based on the Github star number, as well as their versions?} \\*
        \cmidrule{2-2}
         & {(2) Among all NPM packages with project license `MIT' and marked as release, which 5 projects have the highest GitHub fork count?} \\
        \hline
        \multirow[t]{4}{*}{\dst{GITHUB\_REPOS}} & {(1) Among repositories that do not use Python, what proportion of their README.md files include copyright information?} \\*
        \cmidrule{2-2}
         & {(2) Identify the repository in Swift language that contains the most frequently copied non-binary Swift file in the dataset, ensuring that each file is uniquely determined by its ID.} \\*
        \cmidrule{2-2}
         & {(3) How many commit messages are found in repositories that use the Shell programming language and are licensed under Apache-2.0, where each message exists, is shorter than 1,000 characters, and does not begin with `merge', `update', or `test'?} \\*
        \cmidrule{2-2}
         & {(4) List the repository names for the top five GitHub repositories whose main language is not Python, ordered by the highest number of commits.} \\
        \hline
        \multirow[t]{4}{*}{\dst{googlelocal}} & {(1) What are the top 5 businesses located in Los Angeles, California, ranked by highest average rating in descending order?} \\*
        \cmidrule{2-2}
         & {(2) Which massage therapy businesses have an average rating of at least 4.0, and what are their respective average ratings?} \\*
        \cmidrule{2-2}
         & {(3) What are the top 5 businesses that remain open after 6:00 PM on at least one weekday, ranked by highest average rating? Include their names, operating hours, and average ratings.} \\*
        \cmidrule{2-2}
         & {(4) Which 3 businesses received the highest number of reviews with ratings of 4.5 or higher during 2019? Include their names and the count of high-rating reviews.} \\
        \hline
        \multirow[t]{10}{*}{\dst{imdb}} & {(1) Among movies, TV movies, video movies, and video games produced after 1990 that are tagged with at least one keyword, are associated with a US company, have a US internet release date recorded in the 1990s or 2000s, and have a complete and verified cast listing---what is the alphabetically first title, and what type of title is it?} \\*
        \cmidrule{2-2}
         & {(2) Among writers who have a registered pseudonym and are credited on movies associated with a US company, what is the alphabetically first pseudonym and the alphabetically first movie title?} \\*
        \cmidrule{2-2}
         & {(3) Among female actresses credited with a voice role---including general voice, uncredited voice, Japanese dubbed version, or English dubbed version---in movies produced after 2000 that are associated with a US company and have a release date on record, where the actress has a registered alternate name and played a named character, what is the alphabetically first actress name and the alphabetically first movie title?} \\*
        \cmidrule{2-2}
         & {(4) Among cast and crew members whose name starts with `Z' and who are credited in movies tagged with the keyword `character-name-in-title' that are associated with at least one company, what is the alphabetically first name?} \\*
        \cmidrule{2-2}
         & {(5) Among titles of kind `movie' that have a release date and a rating on record and are associated with a US production company, what are the alphabetically first company name, the alphabetically first rating value, and the alphabetically first title?} \\*
        \cmidrule{2-2}
         & {(6) Among female actresses whose name contains `An', who have a registered alternate name, played a named character, and have trivia on record, and who are credited with a voice role in a computer-animated film produced between 2000 and 2010 with a Japan or USA release, associated with a US company---what are the alphabetically first character name, actress name, and title?} \\*
        \cmidrule{2-2}
         & {(7) Among cast and crew members who have a registered pseudonym and are credited in movies tagged with the keyword `character-name-in-title' that are associated with a US company, what are the alphabetically first pseudonym and title?} \\*
        \cmidrule{2-2}
         & {(8) Among female actresses credited with a voice role who have a registered alternate name (with the alternate name belonging to the same person holding the credit) and played a named character, in movies associated with a US company, what are the alphabetically first alternate name, character name, actress name, and title?} \\*
        \cmidrule{2-2}
         & {(9) Among male writers credited in a writing role on Horror or Thriller movies tagged with at least one of `murder', `violence', `blood', `gore', `death', `female-nudity', or `hospital', that have a vote count on record and are associated with a Lionsgate company---what are the alphabetically first genre, vote count, writer name, and title?} \\*
        \cmidrule{2-2}
         & {(10) Among cast members who played a character whose name contains `man' or `Man', in titles of kind `movie' produced after 2000, tagged with at least one of `superhero', `marvel-comics', `based-on-comic', `tv-special', `fight', `violence', `magnet', `web', `claw', or `laser', and with a complete cast listing---what are the alphabetically first cast member name and title?} \\
        \hline
        \multirow[t]{10}{*}{\dst{krama}} & {(1) According to NIFC, among the regions that have a corresponding NIFC Geographic Area Coordination Center (GACC), which GACC accumulated the most human-caused wildfire acres from 2020 to 2023 (inclusive)? Use the latest report (by reported date) within each year. Give the GACC abbreviation.} \\*
        \cmidrule{2-2}
         & {(2) According to NIFC, from 2014 to 2024 (inclusive), which US state has the highest total number of lightning-caused wildfires across all NIFC Geographic Area Coordination Centers whose boundaries intersect with that state?} \\*
        \cmidrule{2-2}
         & {(3) What is the average number of reported identity thefts (as reported in 2024) across all US metropolitan statistical areas with a 2023 population exceeding one million? Estimate each MSA's 2023 population by linearly interpolating between its 2020 census count and 2024 estimate. Round to 4 decimal places.} \\*
        \cmidrule{2-2}
         & {(4) What is the absolute population change (in number of people) from 2020 to 2024 of the metropolitan statistical area with the highest reported identity theft rate per 100K population in the 2024 report?} \\*
        \cmidrule{2-2}
         & {(5) From 2012 to 2018 (inclusive), what is the average water temperature (in $^{\circ}$C) of the beach that had the most water samples failing to meet bacterial water quality requirements? Round to 2 decimal places.} \\*
        \cmidrule{2-2}
         & {(6) From 2008 to 2016 (inclusive), what is the bacterial exceedance rate (as a percentage, rounded to 2 decimal places) of the beach with the lowest average water temperature among all temperature readings collected in that period?} \\*
        \cmidrule{2-2}
         & {(7) Among patients recorded in the UCEC\_CPTAC3\_meta\_table\_V2.1 sheet, what is the average age of patients whose hyperactivated protein is CDK12? Round to 2 decimal places.} \\*
        \cmidrule{2-2}
         & {(8) Among patients recorded in the UCEC\_CPTAC3\_meta\_table\_V2.1 sheet with Serous histologic type who were not excluded from the study, what is the hyperactivated protein of the oldest patient?} \\*
        \cmidrule{2-2}
         & {(9) What is the mean Kp index, according to the 2024 OMNI2 low-resolution hourly data, for the calendar month in 2024 during which Swarm-B recorded its highest mean thermospheric density? Round to 4 decimal places.} \\*
        \cmidrule{2-2}
         & {(10) During the geomagnetic storm period captured in the space-track orbital observations for satellite NORAD 58214, what is the mean BSTAR drag coefficient of satellite NORAD 43180 as recorded in its TLE data? Round to 6 decimal places.} \\
        \hline
         \multirow[t]{3}{*}{\dst{music\_brainz\_20k}} & {(1) How much revenue in USD did Apple Music make from Beyoncé's song `Get Me Bodied' in Canada?} \\*
        \cmidrule{2-2}
         & {(2) Which store earned the most revenue in USD from Brucqe Maginnis' song `Street Hype' across all countries?} \\*
        \cmidrule{2-2}
         & {(3) Which song generated the highest total revenue in USD across all stores and countries?} \\
         \hline
        \multirow[t]{3}{*}{\dst{pancancer\_atlas}} & {(1) For LGG patients, compute the average log10-transformed expression of the IGF2 gene across different histology types. Only include patients with valid IGF2 expression values and histology annotations that are not enclosed in square brackets. Report the final average values with at least four decimal places of precision.} \\*
        \cmidrule{2-2}
         & {(2) Among BRCA patients in the PanCancer Atlas who are alive, which top three histological types show the highest percentage of CDH1 gene mutations?} \\*
        \cmidrule{2-2}
         & {(3) Calculate the chi-square statistic to assess the association between histological types and the presence of CDH1 gene mutations in female BRCA patients from the PanCancer Atlas, excluding categories with marginal totals less than or equal to 10, and only focusing on patients with known histological types and consider only reliable mutation entries.} \\
        \hline
        \multirow[t]{3}{*}{\dst{patents}} & {(1) Identify the CPC technology areas with the highest exponential moving average of patent filings each year (smoothing factor 0.2), and return only the CPC group codes at level 5 whose best year is 2022.} \\*
        \cmidrule{2-2}
         & {(2) Find the CPC technology areas in Germany with the highest exponential moving average of patent filings each year (smoothing factor 0.1) for patents granted in the second half of 2019. Include the full title, CPC group code, and the best year for each CPC group at level 4.} \\*
        \cmidrule{2-2}
         & {(3) Which assignees, excluding UNIV CALIFORNIA itself, have cited patents assigned to UNIV CALIFORNIA, and what are the titles of the primary CPC subclasses associated with these citations? Please provide the name of each citing assignee together with the full title of the CPC subclass.} \\
        \hline
        \multirow[t]{3}{*}{\dst{stockindex}} & {(1) Which stock index in the Asia region has exhibited the highest average intraday volatility since 2020?} \\*
        \cmidrule{2-2}
         & {(2) Among North American stock indices, which indices had more up days than down days in 2018?} \\*
        \cmidrule{2-2}
         & {(3) If an investor had made regular monthly investments in all indices since 2000, which 5 indices would have produced the highest overall returns, and what countries do they belong to?} \\
        \hline
        \multirow[t]{5}{*}{\dst{stockmarket}} & {(1) What was the maximum adjusted closing price in 2020 for The RealReal, Inc.?} \\*
        \cmidrule{2-2}
         & {(2) List all ETF securities listed on NYSE Arca that reached an adjusted closing price above \$200 at any point during 2015, and also report the total number of such ETFs.} \\*
        \cmidrule{2-2}
         & {(3) List all company names on the NASDAQ-listed Market that were financially troubled (delinquent, deficient, or both) and have trading volume in 2008, for each, report its exisiting non-null average daily trading volume in 2008.} \\*
        \cmidrule{2-2}
         & {(4) What are the names (not symbol) of the top 5 non-ETF stocks listed on the New York Stock Exchange (NYSE) that had more up days than down days in 2017? (Up days: closing price > opening price; Down days: closing price < opening price)} \\*
        \cmidrule{2-2}
         & {(5) Which 5 companies listed on the NASDAQ Capital Market had the highest number of days in 2019 where the intraday price range exceeded 20\% of the low price, list the company names please?} \\
        \hline
        \multirow[t]{10}{*}{\dst{usaspending}} & {(1) How many distinct contract awards from the Department of Defense have an award amount greater than \$1,000,000?} \\*
        \cmidrule{2-2}
         & {(2) How many contract awards went to recipients located in California with an award amount greater than \$1,000,000?} \\*
        \cmidrule{2-2}
         & {(3) How many distinct recipient UEIs are associated with contracts in NAICS sector 33?} \\*
        \cmidrule{2-2}
         & {(4) Among awarding agencies that have issued at least 10 contracts, which agency has the highest proportion of contracts with an award amount greater than \$1,000,000? Return the agency name, breaking ties alphabetically.} \\*
        \cmidrule{2-2}
         & {(5) How many distinct 2-digit NAICS sectors are represented among contracts with an award amount of at least \$10,000,000?} \\*
        \cmidrule{2-2}
         & {(6) How many distinct recipients in the recipients database are each associated with more than one UEI?} \\*
        \cmidrule{2-2}
         & {(7) Among contracts with an award amount of at least \$10,000,000, which 2-digit NAICS sector has the most contracts? Return the 2-digit sector code, breaking ties by the smaller code.} \\*
        \cmidrule{2-2}
         & {(8) How many contract awards were made to Lockheed Martin by the Department of Defense with an award amount greater than \$1,000,000?} \\*
        \cmidrule{2-2}
         & {(9) How many contracts have no English-language description and an award amount greater than \$1,000,000?} \\*
        \cmidrule{2-2}
         & {(10) Among the top 10 most active contract recipients by total number of contracts, how many of their contracts have a superseded amount entry on record?} \\
        \hline
        \multirow[t]{7}{*}{\dst{yelp}} & {(1) What is the average rating of all businesses located in Indianapolis, Indiana?} \\*
        \cmidrule{2-2}
         & {(2) Which U.S. state has the highest number of reviews, and what is the average rating of businesses in that state?} \\*
        \cmidrule{2-2}
         & {(3) During 2018, how many businesses that received reviews offered either business parking or bike parking?} \\*
        \cmidrule{2-2}
         & {(4) Which business category has the largest number of businesses that accept credit card payments, and what is its average rating?} \\*
        \cmidrule{2-2}
         & {(5) Which U.S. state has the highest number of businesses that offer WiFi, and what is the average rating for those businesses?} \\*
        \cmidrule{2-2}
         & {(6) Which business received the highest average rating between January 1, 2016 and June 30, 2016, and what category does it belong to? Consider only businesses with at least 5 reviews.} \\*
        \cmidrule{2-2}
         & {(7) Among users who registered on Yelp in 2016, which 5 business categories have received the most total reviews from those users since 2016?} \\
    \end{longtable}
\end{center}
\clearpage
\twocolumn

\subsection{Example Database Configuration}
\label{append-sec:db-config}

The YAML configuration file for the \ds{bookreview} dataset
walked through in \Cref{sec:bench-stats} specifies its
database setup. The agent sees only the logical database
names (\ttt{books\_database}, \ttt{review\_database});
physical paths and connection details are hidden behind the
tools:

\begin{yaml}
db_clients:
  books_database:
    db_type: postgres
    db_name: bookreview_db
    sql_file: query_dataset/books_info.sql
  review_database:
    db_type: sqlite
    db_path: query_dataset/review_query.db
\end{yaml}

\subsection{Workload Properties}
\label{append-sec:workload-prop}
\crdy{To show the mapping between the datasets and the workload properties we observed in formative study, we report in \Cref{tab:workload-prop-queries} which queries in each dataset exhibit P2, ill formatted join keys; P3, unstructured text transformation; and P4, domain knowledge. Each cell lists the relevant query IDs, followed by the number of queries in parentheses. We omit P1, integration across heterogeneous databases, from the table because all 104 queries exhibit it.}

\crdy{\Cref{sec:bench-constr-method} already describes and illustrates the four techniques used to introduce P2 and P3. Here, we name these techniques explicitly and show where each is applied:
\begin{itemize}[nosep, leftmargin=*]
    \item T1: Rename a shared identifier across tables, such that the same entity is identified by columns with different names.
    \item T2: Apply row specific corruptions to join keys, including prefixes, OCR like substitutions, whitespace, and special characters.
    \item T3: Embed values needed to answer a query within unstructured text and remove the original structured column.
    \item T4: Express numeric values in varied free text formats.
\end{itemize}
Some datasets exhibit P2 or P3 ``natively,” or without a synthetic transformation. Specifically, \ds{agnews} natively requires article classification, \ds{music\_brainz\_20k} natively requires entity resolution, and \ds{stockindex} natively requires mapping full stock exchange names to their abbreviations. \Cref{tab:workload-prop-techniques} maps each dataset to the techniques used to introduce P2 and P3, or marks the property as native.
}
\begin{table}[t]
    \crdy{%
    \centering
    \caption{Queries in each dataset that exhibit P2 (ill-formatted join keys), P3 (unstructured text transformation), and P4 (domain knowledge). Each cell lists the query IDs, with the number of such queries in parentheses. P1 (multi-database integration) is omitted because all 104 queries exhibit it.}
    \label{tab:workload-prop-queries}
    \scriptsize
    \setlength{\tabcolsep}{3pt}
    \renewcommand{\arraystretch}{0.92}
    \begin{tabular}{@{}lccc@{}}
    \toprule
    {\bf Dataset} & {\bf P2} & {\bf P3} & {\bf P4} \\
    \midrule
    \ds{agnews} & --- (0) & 1--4 (4) & --- (0) \\
    \ds{bookreview} & 1--3 (3) & 1--3 (3) & --- (0) \\
    \ds{civic\_unstructured} & 1--10 (10) & 1--10 (10) & 3--6, 10 (5) \\
    \ds{crmarenapro} & 1--13 (13) & 1--3, 6--7, 11 (6) & 1--3, 6--8, 10--13 (10) \\
    \ds{cve} & 1--10 (10) & 1--10 (10) & 1--10 (10) \\
    \ds{deps\_dev\_v1} & --- (0) & 1--2 (2) & 1--2 (2) \\
    \ds{github\_repos} & --- (0) & 1--4 (4) & 1--4 (4) \\
    \ds{googlelocal} & --- (0) & 1--4 (4) & --- (0) \\
    \ds{imdb} & 1--10 (10) & --- (0) & 6, 9--10 (3) \\
    \ds{krama} & 1--10 (10) & 1--10 (10) & 1--2, 5--10 (8) \\
    \ds{music\_brainz\_20k} & --- (0) & 1--3 (3) & --- (0) \\
    \ds{pancancer\_atlas} & --- (0) & 1--3 (3) & 1--3 (3) \\
    \ds{patents} & --- (0) & 1--3 (3) & 1--3 (3) \\
    \ds{stockindex} & 1--3 (3) & 1--3 (3) & 1--3 (3) \\
    \ds{stockmarket} & --- (0) & 1--5 (5) & 1--5 (5) \\
    \ds{usaspending} & 1--10 (10) & 1--10 (10) & 1--3, 5--8, 10 (8) \\
    \ds{yelp} & 1--7 (7) & 1--7 (7) & --- (0) \\
    \midrule
    {\bf Total} & {\bf 76/104} & {\bf 87/104} & {\bf 64/104} \\
    \bottomrule
    \end{tabular}%
    }
\end{table}

\begin{table}[t]
    \crdy{%
    \centering
    \caption{Techniques used to introduce P2 (ill-formatted join keys) and P3 (unstructured text transformation) in each dataset. ``native'' marks datasets that exhibit the property without any synthetic transformation.}
    \label{tab:workload-prop-techniques}
    \scriptsize
    \setlength{\tabcolsep}{3pt}
    \renewcommand{\arraystretch}{0.92}
    \begin{tabular}{@{}lcc@{}}
    \toprule
    {\bf Dataset} & {\bf P2 technique(s)} & {\bf P3 technique(s)} \\
    \midrule
    \ds{agnews} & --- & native \\
    \ds{bookreview} & T1 & T3 \\
    \ds{civic\_unstructured} & T1, T2 & T3 \\
    \ds{crmarenapro} & T2 & T3, T4 \\
    \ds{cve} & T2 & T3, T4 \\
    \ds{deps\_dev\_v1} & --- & T3 \\
    \ds{github\_repos} & --- & T3 \\
    \ds{googlelocal} & --- & T3 \\
    \ds{imdb} & T2 & --- \\
    \ds{krama} & T1, T2 & T3 \\
    \ds{music\_brainz\_20k} & --- & native \\
    \ds{pancancer\_atlas} & --- & T3 \\
    \ds{patents} & --- & T3 \\
    \ds{stockindex} & native & T3, T4 \\
    \ds{stockmarket} & --- & T3 \\
    \ds{usaspending} & T2 & T4 \\
    \ds{yelp} & T1 & T3 \\
    \bottomrule
    \end{tabular}%
    }
\end{table}

\section{Tools and Prompts}
\label{append-sec:prompt-templates}

\autoref{tab:tool_description} shows the four tools provided to Gemini-3.1-Pro (ReAct).

\begin{table*}[t]
    \centering
    \caption{Description of tools given to Gemini-3.1-Pro (ReAct).}
    \label{tab:tool_description}
    \scriptsize
    \setlength{\tabcolsep}{3pt}
    \renewcommand{\arraystretch}{0.92}
    \begin{tabular}{@{}l p{0.1\textwidth} p{0.32\textwidth} p{0.41\textwidth}@{}}
        \toprule
         {\bf Tool} & {\bf Argument(s)} & {\bf Description} & {\bf Example} \\
         \midrule
         \texttt{\scriptsize list\_db} & \texttt{\scriptsize db\_name} &  Returns the names of all tables in the database specified by \texttt{\scriptsize db\_name}, or an error if it does not exist. & \texttt{\scriptsize list\_db(\textquotesingle books\textquotesingle)} returns the names of the tables in the \texttt{\scriptsize books} database: \texttt{\scriptsize [books\_info, authors\_info]}. \\
         \hline
         \texttt{\scriptsize query\_db} & \texttt{\scriptsize db\_name}, \texttt{\scriptsize query}  & Verifies that \texttt{\scriptsize query}, written in SQL or Mongo, is read-only and, if valid, executes it on the database specified by \texttt{\scriptsize db\_name}, returning the query result or an error if execution fails. & \texttt{\scriptsize query\_db(\textquotesingle books\textquotesingle, \textquotesingle SELECT title, details FROM books\_info;\textquotesingle)} executes the specified query on the \texttt{\scriptsize books} database and returns the tuples from the \texttt{\scriptsize books\_info} table with the specified columns, stored in the variable \texttt{\scriptsize var\_call\_123}.\\
         \hline
         \texttt{\scriptsize execute\_python} & \texttt{\scriptsize code} &  Executes triple-quoted Python \texttt{\scriptsize code} via \texttt{\scriptsize exec(code, env)} in a Docker environment with Python~3.12 and two popular data-processing libraries, Pandas and Pyarrow,  preinstalled, with a 600s timeout. Here, \texttt{\scriptsize env} is a dictionary that maps variable names to prior tool call results. The execution result must be JSON-serializable and printed with the prefix line \texttt{\scriptsize \_\_RESULT\_\_:} to be correctly parsed. &  \texttt{\scriptsize execute\_python(\textquotedbl \textquotedbl \textquotedbl import pandas as pd\newline df = pd.DataFrame(var\_call\_123)\newline res = df[df[\textquotesingle details\textquotesingle].str.contains(\allowbreak\textquotesingle English\textquotesingle, regex=False)].copy()\newline print(\textquotesingle\_\_RESULT\_\_:\textquotesingle)\newline print(json.dumps(\newline res.to\_dict(orient=\textquotesingle records\textquotesingle))\textquotedbl \textquotedbl \textquotedbl)}
         returns the books written in English extracting language information from the free-text \texttt{\scriptsize details} column.\\
         \hline
         \texttt{\scriptsize return\_answer} & \texttt{\scriptsize answer} & Returns \texttt{\scriptsize answer} and terminates agent execution. & The agent terminates and returns the answer using \texttt{\scriptsize return\_answer(\textquotesingle The books written in English are Fire Cracker, $\ldots$\textquotesingle)}.\\
         \bottomrule
    \end{tabular}
\end{table*}

Below is the system prompt for Gemini-3.1-Pro\footnote{The system prompts for all models are available at: \techreport{\url{https://drive.google.com/drive/folders/14SiSz7CVnQA58YOAzFzNjoUu3Wnmfe5t?usp=drive_link}}\papertext{{\color{red!70!black}[URL redacted for submission]}}}.

\begin{lstlisting}[language=text,style=plaintextstyle]
You are a data analysis agent. Use only the tools listed below to answer the user's query, based on the provided DATABASE DESCRIPTION for logical database names and their types (SQL or MongoDB), and the results of previous tool calls.

TOOLS (system will execute):
- query_db: run a SQL or Mongo query. Returns a list of JSON-serializable records or an error string.
  Required args: {"db_name": "<logical_db_name>", "query": "<SQL or Mongo query>"}
- list_db: list tables or collections for a given database.
  Required args: {"db_name": "<logical_db_name>"}
- execute_python: run Python code.
  Required args: {"code": "<python_code>"}
- return_answer: finish and return the final answer (plain text).
  Required args: {"answer": "<final plain-text answer>"}

INSTRUCTIONS:
1. After each tool call, its result will be stored in a storage under a key named after the tool call id (you will be told the key name). The next message will include both the result (or a preview if it's large) and the storage key name.
2. Inside execute_python code you may read storage entries using the provided key names, e.g., if the tool call id is 'call-1', you can access its result via `locals()['var_call-1']` in your code.
3. You cannot modify or reassign those storage-provided variables; you may read them and create new variables as needed.
4. If a tool result is large, the next message will include a preview (first 10000 characters) and the storage entry will be the .json file path (a string) where the full result is stored. To access the full result, your execute_python code must open and read that .json file.

KEY RULES (must follow exactly):
1. Always use tool calls. Do not output plain text, explanations, or reasoning.
2. Include all required arguments for the tool you call.
3. For query_db, always specify db_name and query. Refer the DATABASE DESCRIPTION for db_name and query format.
4. For PostgreSQL, wrap mixed-case or uppercase column names in double quotes.
5. For list_db, refer the DATABASE DESCRIPTION to specify db_name.
6. Use execute_python for data processing as needed. 
7. When using execute_python, your code will be quoted by triple double-quotes and passed as a string to `exec(...)` for execution in a Python 3.12 environment with only pandas and pyarrow installed. So you must ensure your code is compatible with this execution method. For exampe, do not use triple double-quotes in your code, as they may interfere with parsing. Do not use non-built-in or non-installed packages.
8. When using execute_python, your code must print the result at the end exactly as shown in the PRINT FORMAT section below. The printed result must be a string that can be successfully parsed by json.loads() without errors.

PRINT FORMAT (must match exactly):
----BEGIN PRINT FORMAT----
print("__RESULT__:")
print(your_json_serializable_string_here)
----END PRINT FORMAT----

For simple types (int, float, str, bool, None), you may use json.dumps() to produce a valid JSON string.
For complex or non-JSON-serializable types, you must convert them into JSON-compatible forms before printing.
For lists or dictionaries, you must ensure that all nested elements are also converted into JSON-serializable types.
9. Return the final answer only via a single return_answer tool call. Do not include extra text, explanation, or formatting.

EXAMPLES:
- query_db:
{"tool": "query_db", "args": {"db_name": "some_db_name", "query": "SELECT * FROM some_table LIMIT some_limit;"}} # for SQL databases
{"tool": "query_db", "args": {"db_name": "some_db_name", "query": "{"collection": "some_collection", "filter": {some_filter}, "projection": {some_projection}, "limit": some_limit}"}} # for MongoDB databases

- list_db:
{"tool": "list_db", "args": {"db_name": "some_db_name"}}

- execute_python:
{"tool": "execute_python", "args": {"code": "import pandas as pd
# rl1 and rl2 are the keys of two JSON-serializable record lists in storage
df1 = pd.DataFrame(rl1)
df2 = pd.DataFrame(rl2)
result = pd.merge(df1, df2, on='id').head(10).to_json(orient='records')
print('__RESULT__:')
print(result)"}}

- return_answer:
{"tool": "return_answer", "args": {"answer": "...final plain-text answer..."}}

If you cannot proceed, call return_answer with a short explanatory message.

Do not output explanations, reasoning, or any natural language outside of the required tool calls.
\end{lstlisting}

\section{Shared Harness Evaluation Results}
\label{append-sec:shared-harness-results}

\crdy{To further distinguish different models, we evaluate them using a shared ReAct harness (as described in \Cref{sec:exp-setup}) with GPT-5.2, Claude Opus 4.6, and Gemini 3.1 Pro. Due to budget limits, we evaluated about half of the queries; 54 queries across 12 datasets (selecting the least expensive queries), with five trials per model per query. Results are reported in \Cref{tab:shared-harness-per-dataset}.}

\begin{table}[t]
    \crdy{%
    \centering
    \caption{Per-dataset pass@1 (P@1) and pass@5 (P@5) under the shared ReAct harness, over 54 queries across 12 datasets with five trials per model per query. \best{Green}: highest per dataset and metric; \zv{}: zero score.}
    \label{tab:shared-harness-per-dataset}
    \scriptsize
    \setlength{\tabcolsep}{3pt}
    \renewcommand{\arraystretch}{0.92}
    \begin{tabular}{@{}l rr rr rr@{}}
    \toprule
    & \multicolumn{2}{c}{\bf GPT-5.2} & \multicolumn{2}{c}{\bf Claude-Opus-4.6} & \multicolumn{2}{c}{\bf Gemini-3.1-Pro} \\
    \cmidrule(lr){2-3}\cmidrule(lr){4-5}\cmidrule(lr){6-7}
    {\bf Dataset} & {\bf P@1} & {\bf P@5} & {\bf P@1} & {\bf P@5} & {\bf P@1} & {\bf P@5} \\
    \midrule
    \ds{agnews}             & \zv         & \zv         & \best{0.30} & \best{0.50} & 0.25        & \best{0.50} \\
    \ds{bookreview}         & 0.53        & 0.67        & \best{1.00} & \best{1.00} & 0.93        & \best{1.00} \\
    \ds{crmarenapro}        & 0.52        & 0.77        & \best{0.79} & \best{0.92} & 0.75        & 0.85        \\
    \ds{deps\_dev\_v1}      & \zv         & \zv         & 0.40        & \best{0.50} & \best{0.50} & \best{0.50} \\
    \ds{github\_repos}      & 0.15        & 0.25        & 0.35        & \best{0.50} & \best{0.40} & \best{0.50} \\
    \ds{googlelocal}        & 0.30        & 0.50        & 0.75        & 0.75        & \best{0.80} & \best{1.00} \\
    \ds{music\_brainz\_20k} & 0.07        & 0.33        & 0.40        & 0.67        & \best{0.73} & \best{1.00} \\
    \ds{pancancer\_atlas}   & 0.40        & \best{0.67} & 0.47        & \best{0.67} & \best{0.53} & \best{0.67} \\
    \ds{patents}            & \zv         & \zv         & \zv         & \zv         & \zv         & \zv         \\
    \ds{stockindex}         & \best{0.93} & \best{1.00} & 0.73        & \best{1.00} & \best{0.93} & \best{1.00} \\
    \ds{stockmarket}        & 0.32        & 0.80        & 0.72        & 0.80        & \best{0.80} & \best{1.00} \\
    \ds{yelp}               & 0.20        & 0.29        & \best{0.66} & \best{1.00} & 0.63        & \best{1.00} \\
    \midrule
    {\bf Stratified avg}    & 0.29        & 0.44        & 0.55        & 0.69        & \best{0.60} & \best{0.75} \\
    \bottomrule
    \end{tabular}%
    }
\end{table}

\section{Ablation Results}
\label{append-sec:abl_results}

\autoref{tab:acc_per_dataset_nohints} and \autoref{tab:acc_per_dataset_nodesc_nohints} report per-dataset average pass@1 accuracy when agents run without hints (``w/o hints'') and without both hints and description files (``w/o desc + hints''), respectively.
\autoref{tab:cost_nohints} and \autoref{tab:cost_nodescrip} report per-dataset cost (in USD) of ``w/o hints'' and ``w/o desc + hints'', respectively.
\autoref{tab:exploration} reports the number of exploratory tool calls per trial in each setting.

\begin{table}[t]
    \centering
    \caption{Number of exploratory tool calls (e.g., schema
    lookups or SELECT * LIMIT 5) per trial, with share of
    all tool calls in parentheses, averaged over one trial
    per query. A Claude Code subagent classifies each call.}
    \label{tab:exploration}
    \scriptsize
    \setlength{\tabcolsep}{3pt}
    \renewcommand{\arraystretch}{0.92}
    \begin{tabular}{@{}l c c c@{}}
      \toprule
      {\bf Setting} & {\bf Claude-Sonnet-4.6} & {\bf GPT-5.5} & {\bf Gemini-3.1-Pro} \\
      \midrule
      full info          & 14.5 (39.2\%) & 7.1 (50.9\%)  & 6.9 (21.8\%) \\
      w/o hints          & 15.0 (45.8\%) & 6.2 (45.0\%)  & 7.3 (23.6\%) \\
      w/o desc + hints   & 21.1 (55.1\%) & 10.4 (60.6\%) & 14.5 (40.9\%) \\
      \bottomrule
    \end{tabular}
\end{table}

\begin{table}[t]
    \centering
    \caption{Pass@1 score by dataset when agents run without hints (``w/o hints''), with the average across all 17 datasets.
    One dataset remains completely unsolved by every agent.
    \best{Green}: highest per dataset; \zv{}: zero Pass@1; \blank{Gray}: partial results, where some queries in the dataset have not yet been evaluated.}
    \label{tab:acc_per_dataset_nohints}
    \scriptsize
    \setlength{\tabcolsep}{3pt}
    \renewcommand{\arraystretch}{0.92}
    \begin{tabular}{@{}lrrr@{}
    }
    \toprule
    {\bf Dataset} & {\bf Claude-Sonnet-4.6} & {\bf GPT-5.5} & {\bf Gemini-3.1-Pro} \\
    \midrule
    \ds{agnews} & \blank{0.35} & 0.45 & 0.30 \\
    \ds{bookreview} & 0.80 & \best{1.00} & \best{1.00} \\
    \ds{civic\_unstructured} & \best{0.50} & 0.48 & 0.48 \\
    \ds{crmarenapro} & 0.69 & 0.75 & \best{0.77} \\
    \ds{cve} & \blank{0} & \best{0.18} & 0.08 \\
    \ds{deps\_dev\_v1} & 0.30 & \best{0.40} & \best{0.40} \\
    \ds{github\_repos} & \best{0.45} & 0.35 & 0.30 \\
    \ds{googlelocal} & 0.70 & \best{1.00} & 0.70 \\
    \ds{imdb} & 0.58 & \best{0.64} & 0.36 \\
    \ds{krama} & 0.56 & \best{0.74} & 0.16 \\
    \ds{music\_brainz\_20k} & \zv & \best{0.27} & \best{0.27} \\
    \ds{pancancer\_atlas} & \best{0.67} & 0.60 & 0.60 \\
    \ds{patents} & \zv & \zv & \zv \\
    \ds{stockindex} & 0.87 & 0.80 & \best{1.00} \\
    \ds{stockmarket} & 0.92 & \best{0.96} & 0.72 \\
    \ds{usaspending} & 0.40 & 0.38 & \best{0.42} \\
    \ds{yelp} & 0.66 & \best{0.91} & 0.66 \\
    \midrule
    {\bf Average} & 0.50 & \best{0.58} & 0.48 \\
    \bottomrule
    \end{tabular}
\end{table}

\begin{table}[t]
    \centering
    \caption{Pass@1 score by dataset when agents run without both hints and description files (``w/o desc + hints''), with the average across all 17 datasets.
    One dataset remains completely unsolved by every agent.
    \best{Green}: highest per dataset; \zv{}: zero Pass@1; \blank{Gray}: partial results, where some queries in the dataset have not yet been evaluated.}
    \label{tab:acc_per_dataset_nodesc_nohints}
    \scriptsize
    \setlength{\tabcolsep}{3pt}
    \renewcommand{\arraystretch}{0.92}
    \begin{tabular}{
    @{}lrrr@{}
    }
    \toprule
    {\bf Dataset} & {\bf Claude-Sonnet-4.6} & {\bf GPT-5.5} & {\bf Gemini-3.1-Pro} \\
    \midrule
    \ds{agnews} & \blank{0.10} & 0.30 & 0.30 \\
    \ds{bookreview} & 0.73 & \best{1.00} & \best{1.00} \\
    \ds{civic\_unstructured} & 0.44 & \best{0.52} & 0.32 \\
    \ds{crmarenapro} & 0.68 & \best{0.82} & 0.80 \\
    \ds{cve} & \blank{0.06} & \best{0.22} & 0.12 \\
    \ds{deps\_dev\_v1} & 0.20 & \best{0.50} & 0.40 \\
    \ds{github\_repos} & \best{0.50} & 0.30 & 0.40 \\
    \ds{googlelocal} & 0.75 & \best{0.80} & 0.70 \\
    \ds{imdb} & \blank{0.40} & \best{0.56} & 0.42 \\
    \ds{krama} & \blank{0.16} & \best{0.68} & 0.14 \\
    \ds{music\_brainz\_20k} & 0.07 & \best{0.27} & 0.20 \\
    \ds{pancancer\_atlas} & \best{0.67} & 0.53 & 0.47 \\
    \ds{patents} & \best{0.07} & \zv & \zv \\
    \ds{stockindex} & \best{0.93} & \best{0.93} & \best{0.93} \\
    \ds{stockmarket} & \best{0.92} & 0.84 & 0.72 \\
    \ds{usaspending} & \blank{0} & 0.30 & \best{0.32} \\
    \ds{yelp} & 0.71 & \best{0.83} & 0.60 \\
    \midrule
    {\bf Average} & 0.43 & \best{0.55} & 0.47 \\
    \bottomrule
    \end{tabular}
\end{table}

\begin{table}[t]
    \centering
    \caption{Total cost (USD) per dataset when agents run without hints (``w/o hints''). \best{Green}: cheapest per dataset; \blank{Gray}: estimated total cost for datasets with partial results.}
    \label{tab:cost_nohints}
    \scriptsize
    \setlength{\tabcolsep}{3pt}
    \renewcommand{\arraystretch}{0.92}
    \begin{tabular}{
    @{}lrrr@{}
    }
    \toprule
    {\bf Dataset} & {\bf Claude-Sonnet-4.6} & {\bf GPT-5.5} & {\bf Gemini-3.1-Pro} \\
    \midrule
    \ds{agnews} & \blank{23.8} & 34.7 & 10.7 \\
    \ds{bookreview} & 3.7 & 3.0 & 1.8 \\
    \ds{civic\_unstructured} & 108.6 & 59.9 & 26.9 \\
    \ds{crmarenapro} & 26.1 & 23.3 & 16.9 \\
    \ds{cve} & \blank{273.4} & 129.0 & 94.7 \\
    \ds{deps\_dev\_v1} & 3.5 & 5.3 & 2.8 \\
    \ds{github\_repos} & 4.4 & 9.7 & 4.7 \\
    \ds{googlelocal} & 3.8 & 3.9 & 1.6 \\
    \ds{imdb} & 46.8 & 80.6 & 23.9 \\
    \ds{krama} & 90.9 & 58.1 & 49.8 \\
    \ds{music\_brainz\_20k} & 0.9 & 1.8 & 1.3 \\
    \ds{pancancer\_atlas} & 5.6 & 5.9 & 3.0 \\
    \ds{patents} & 8.5 & 11.9 & 4.9 \\
    \ds{stockindex} & 2.7 & 3.2 & 2.6 \\
    \ds{stockmarket} & 3.4 & 3.3 & 9.2 \\
    \ds{usaspending} & 54.0 & 36.9 & 16.4 \\
    \ds{yelp} & 10.2 & 11.8 & 8.3 \\
    \midrule
    {\bf Total} & 670.4 & 482.5 & 279.5 \\
    \bottomrule
    \end{tabular}
\end{table}

\begin{table}[t]
    \centering
    \caption{Total cost (USD) per dataset when agents run without both hints and description files (``w/o desc + hints''). \best{Green}: cheapest per dataset; \blank{Gray}: estimated total cost for datasets with partial results.}
    \label{tab:cost_nodescrip}
    \scriptsize
    \setlength{\tabcolsep}{3pt}
    \renewcommand{\arraystretch}{0.92}
    \begin{tabular}{
    @{}lrrr@{}
    }
    \toprule
    {\bf Dataset} & {\bf Claude-Sonnet-4.6} & {\bf GPT-5.5} & {\bf Gemini-3.1-Pro} \\
    \midrule
    \ds{agnews} & \blank{26.4} & 46.7 & 13.4 \\
    \ds{bookreview} & 3.8 & 4.4 & 3.3 \\
    \ds{civic\_unstructured} & 111.0 & 52.9 & 31.5 \\
    \ds{crmarenapro} & 29.7 & 32.4 & 16.7 \\
    \ds{cve} & \blank{204.6} & 79.4 & 69.2 \\
    \ds{deps\_dev\_v1} & 3.9 & 9.4 & 3.4 \\
    \ds{github\_repos} & 6.9 & 15.0 & 7.2 \\
    \ds{googlelocal} & 5.2 & 4.7 & 2.0 \\
    \ds{imdb} & \blank{62.4} & 59.0 & 25.4 \\
    \ds{krama} & \blank{107.2} & 62.2 & 70.9 \\
    \ds{music\_brainz\_20k} & 1.3 & 2.7 & 0.7 \\
    \ds{pancancer\_atlas} & 6.0 & 7.4 & 3.0 \\
    \ds{patents} & 11.9 & 14.7 & 6.1 \\
    \ds{stockindex} & 3.1 & 4.7 & 3.2 \\
    \ds{stockmarket} & 5.6 & 6.1 & 15.6 \\
    \ds{usaspending} & \blank{84.7} & 31.1 & 19.8 \\
    \ds{yelp} & 12.3 & 20.3 & 7.9 \\
    \midrule
    {\bf Total} & 686.0 & 453.0 & 299.2 \\
    \bottomrule
    \end{tabular}
\end{table}

\section{Failure Description and Examples}

\subsection{Failure-Mode Frequency Plots}
\label{append-sec:fm-freq-plots}

\begin{figure}[!ht]
    \centering
    \includegraphics[width=0.8\linewidth]{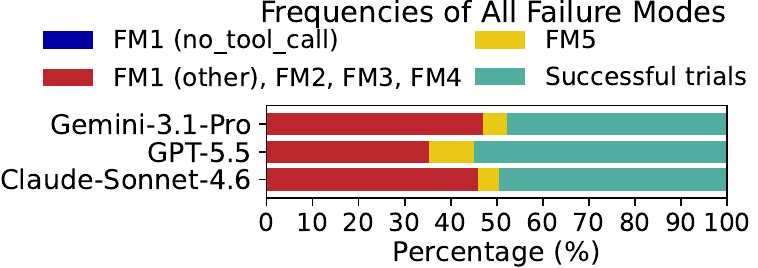}
\caption{Failure-mode breakdown across all 1{,}560 trajectories. Runtime errors (\smallfmbadge{FM5}) are rare; \smallfmbadge{FM1\,(no\_tool\_call)} only occurs in one Gemini-3.1-Pro trial.}
\label{fig:fm-freq}
\end{figure}
\begin{figure}[!ht]
    \centering
    \includegraphics[width=0.8\linewidth]{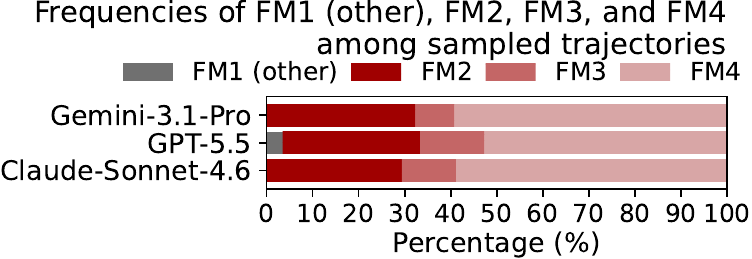}
\caption{Decomposition of the LLM-judged red bar in \Cref{fig:fm-freq} over 196 trajectories. \smallfmbadge{FM4} and \smallfmbadge{FM2} account for 90\% of failures.}
\label{fig:fm234-freq}
\end{figure}
\FloatBarrier

\subsection{FM Annotation Prompt}
\label{append-sec:fm-annot-prompt}

Below is the prompt template for GPT-5 to annotate \fmbadge{FM1 (other)}, \fmbadge{FM2}, \fmbadge{FM3}, and \fmbadge{FM4} for a failed trajectory.
\begin{lstlisting}[language=text,style=plaintextstyle]
You are given a trace of a FAILED task executed by a data agent, along with the task query and the ground-truth answer. Your task is to diagnose WHY the agent's final answer does not match the ground truth using the failure modes FM1-4 defined below.

The agent had access to the following tools:
- list_db: list tables in a database
- query_db: execute SQL queries
- execute_python: run Python code for data processing
- return_answer: terminate and return the final answer

The trace records the complete interaction between the data agent and the available tools. That is, the trace contains:
- Exploratory tool calls (trial-and-error)
- Tool calls that directly contributed to the final answer

## IMPORTANT INSTRUCTIONS:
1. First, identify ONLY the tool calls whose outputs were used (directly or indirectly) to produce the final answer. Ignore abandoned, failed, or exploratory tool calls that did not affect the final output.
2. Base your failure analysis ONLY on those contributing tool calls.
3. You must mark at least one failure mode as "yes". By default, select exactly one failure mode-the most specific one that best explains why the final answer is incorrect. Select multiple failure modes only if the failure cannot be adequately explained by any single mode alone (i.e., each selected mode captures a distinct, necessary cause visible in the contributing tool calls).
4. Only mark a failure mode as "yes" if you can point to a concrete example in the trace.
5. Provide a single-sentence summary explaining the failure, explicitly referencing the trace behavior.

## FAILURE MODELS
FM1 - Fails Before Planning
Definition:
- The agent does not attempt to solve the query.
Example:
- The agent does not issue any tool calls.

FM2 - Incorrect Plan
Definition:
- The agent attempts to solve the user's query, but the PLAN (i.e., the logical structure) of the solution is wrong. That is, even if all steps were executed perfectly (e.g., correct data selection, correct implementation, no execution errors), the plan cannot produce the ground-truth answer.
Examples:
- Missing operations specified or implied by the user.
- Adding constraints not requested by the user.
- Stops early and returns an answer before all requirements of the query are completed.

FM3 - Correct Plan, Wrong Data Selection
Definition:
- The agent follows a theoretically correct plan for answering the user's query, but selects incorrect data sources in its implementation, such that the required information exists but is retrieved from the wrong database, table, collection, column, or field.
Example:
- Using an incorrect column in a selection or filtering condition (e.g., a WHERE clause references a column that does not represent the queried attribute, even though the correct column exists elsewhere in the schema).
- Querying or joining a table that does not represent the queried entity, despite the correct table exists elsewhere in the schema.

FM4 - Correct Plan and Data Selection, Incorrect Implementation
Definition:
The agent the correct plan and selects the correct tables and columns, but implements the computation incorrectly.
Examples:
- Arithmetic errors, such as computing an aggregate with an incorrect formula (e.g., dividing by the number of unique entities rather than the total number of contributing entities or records when calculating an overall average).
- Incorrect regular expressions or parsing rules, such that extracted values include unintended matches or miss required values (e.g., patterns that capture unrelated numbers or fail to match the intended tokens).
- Parsing errors, such as using unsupported syntax, improper escaping, or assumptions about the execution environment that cause incorrect behavior or failed execution.
- Incorrect join implementation, where the agent applies an invalid join condition or fails to normalize or transform identifiers before joining, even though the trace indicates awareness that normalization or alignment is required.

## QUERY
{{uesr_query}}

## GROUND-TRUTH ANSWER
{{ground_truth_answer}}

## FAILED TRACE
{{failed_trajectory}}

## OUTPUT FORMAT
Your output MUST follow this exact format, starting after @@ and ending before @@:

@@
A. Freeform text summary of the failure:
<one sentence>
B. Failure modes encountered:
FM1: <yes or no>
FM2: <yes or no>
FM3: <yes or no>
FM4: <yes or no>
@@
\end{lstlisting}

\subsection{Examples of Failed Trajectories}
\label{append-sec:fm-examples}
Examples in this section are drawn from the bookreview dataset, with the following description and hints:
\begin{lstlisting}[language=text,style=plaintextstyle]
DATABASE DESCRIPTION:
You are working with two databases to solve this query.

Here are the descriptions of these two databases:

1. books_database
   - This database is stored in a PostgreSQL database and contains Amazon book information including descriptions, price, details, title, etc. up to 2023.
   - This database consists of one table:
    - books_info:
      - Fields:
        - title (str): Book title
        - subtitle (str): Book subtitle
        - author (str): Book author(s)
        - rating_number (int): Total number of ratings received
        - features (str): Book features (stored as string representation of list/dict)
        - description (str): Book description (stored as string representation of list/dict)
        - price (float): Book price
        - store (str): Store information
        - categories (str): Book categories (stored as string representation of list/dict)
        - details (str): Additional book details
        - book_id (str): Unique book identifier

2. review_database
   - This database is stored in a SQLite database and contains Amazon book review information including ratings, text, helpfulness votes, etc. up to 2023.
   - This database consists of one table:
    - review
      - Fields:
        - rating (float): Rating given by reviewer (1.0-5.0 scale)
        - title (str): Review Title
        - text (str): Review text content
        - purchase_id (str): Unique identifier linking to book_id in books_info table in books_database
        - review_time (str): Timestamp when review was posted
        - helpful_vote (int): Number of helpful votes received
        - verified_purchase (bool): Whether purchase was verified
\end{lstlisting}
\begin{lstlisting}[language=text,style=plaintextstyle]
HINTS: 
- In books_info, the "description", "categories", and "features" content appears to be in list or dictionary format, but they are actually stored as strings in the .sql file.
- The fields "book_id" in books_info and "purchase_id" in review refer to the same book entities across different tables. While the field names are not exact matches, they can be joined using a fuzzy join approach to solve the query.
- For some queries, you could get needed information from "categories" or "details" columns in books_info.
\end{lstlisting}
There are three queries under bookreview, denoted as Q1, Q2, and Q3. We list the queries and their corresponding ground-truth answers A1, A2, and A3 below:
\begin{enumerate}[leftmargin=0pt, itemindent=*,
                  labelsep=4pt, widest=iii, align=left, nosep]
\item [{\bf (Q1)}] {\em Which decade of publication (e.g., 1980s) has the highest average rating among decades with at least 10 distinct books that have been rated? Return the decade with the highest average rating.}
\item [{\bf (A1)}] 2020
\item [{\bf (Q2)}] {\em Which English-language books in the 'Literature \& Fiction' category have a perfect average rating of 5.0? Return all matching books.}
\item [{\bf (A2)}] The Sludge; 
Something That Feels Like Truth (Switchgrass Books); 
Kennebago Moments;
Hollywood Confessions: Hollywood Headlines Book \#3 (Hollywood Headlines Mysteries);
Forged in Blood (Freehold);
Local Honey;
"Exits, Desires, \& Slow Fires";
Fire Cracker;
Reunion: The Children of Lauderdale Park;
Childe Harold of Dysna;
The Prophet: With Original 1923 Illustrations by the Author;
Knowing When To Die: Uncollected Stories;
Liza of Lambeth;
Child Of The King A Journey of Hope Book 1: Earthly Story With A Heavenly Message;
The Melancholy Strumpet Master;
\item [{\bf (Q3)}] {\em Which books categorized as 'Children's Books' have received an average rating of at least 4.5 based on reviews from 2020 onwards?}
\item [{\bf (A3)}]Around the World Mazes;
Behind the Wheel (Choose Your Own Adventure \#35)(Paperback/Revised);
Benny Goes To The Moon: The great new book from Top Children's entertainer Gerry Ogilvie (1);
"Cheer Up, Ben Franklin! (Young Historians)";
Favorite Thorton W. Burgess Stories: 6 Books;
Egypt (Enchantment of the World);
"Pokemon: Sun \& Moon, Vol. 8 (8)";
The Library Book;
LunaLu the Llamacorn;
Monstrous Stories \#4: The Day the Mice Stood Still;
The Old Man and the Pirate Princess;
Trouble in the CTC!: The Terra Prime Adventures Book 2;
"Clark the Shark: Tooth Trouble, No. 1";
Cleo Porter and the Body Electric;
\end{enumerate}

For brevity, we analyze only the failure modes of selected example trajectories and omit the full interaction traces\footnote{Complete trajectories are available at \techreport{\url{https://drive.google.com/drive/folders/1FES67CWgfOaR-1FRcqoIeyvNaUV4NFZ6?usp=drive_link}}\papertext{{\color{red!70!black}[URL redacted for submission]}}. In these example trajectories, for presentation clarity, we omit system and user prompts, and truncate large tool results to 300 characters.}.

\subsubsection{\fmbadge{FM2}: Incorrect Plan for Averaging}

\topic{Query}
Q1

\topic{Analysis}
This trajectory fails 
due to an incorrect averaging plan: 
it computes decade-level averages by averaging per-book average ratings,
whereas the query requires directly averaging all ratings within each decade.
Specifically,
the final answer in Line 423 
derives from the \ttt{execute\_python} tool call 
in Line 368, 
which uses results from prior \ttt{query\_db} calls 
that retrieve records 
from \texttt{books} (Line 48; \ttt{\small var\_call\_mC9eh9kdqR7TFrzmoKhf7oa0}) 
and \texttt{reviews} (Line 36; \ttt{\small var\_call\_clwW1HpxqxlCKDXJvn9Iim9W}). 
The agent first computes per-book average ratings---treating each \ttt{purchase\_id} 
as a book identifier---when processing review records (Line 38). 
After merging the retrieved \texttt{books} and \texttt{reviews} tables in Python (Line 396), 
it computes decade-level averages by averaging these per-book averages (Line 401), 
resulting in an incorrect aggregation.

\subsubsection{\fmbadge{FM2}: Missing Operations}

\topic{Query}
Q2

\topic{Analysis}
This trajectory fails 
by missing a required constraint:
the selected books must have an average rating of exactly 5.0.
Specifically,
the final answer in Line 591
is produced by the \ttt{execute\_python} tool call in Line 561, 
which depends on a prior \ttt{execute\_python} call in Line 525, 
itself derived from a \ttt{query\_db} call in Line 113. 
All three calls contributing to the final answer ignore the constraint that the average rating must equal 5.0.

\subsubsection{\fmbadge{FM2}: Adding Operations}

\topic{Query}
Q3

\topic{Analysis}
This trajectory fails by introducing unwarranted operations: 
it adds \texttt{LIMIT 200} when retrieving review records (Line 38) 
and \texttt{LIMIT 500} when retrieving book records (Line 50), 
while the query does not require any such limits.

\subsubsection{\fmbadge{FM3}: Incorrect Column Selection}

\topic{Query}
Q2

\topic{Analysis}
This trajectory 
fails by selecting an incorrect column. 
To filter books written in English, 
the agent relies on columns other than the correct one (i.e., \ttt{details}) 
in the \ttt{books\_info} table, 
although the database description already indicates \ttt{details} as the correct column (Line 39).
Specifically, 
in Line 125 of the trajectory below, 
the agent attempts to search for ``English'' 
using \ttt{details}, \ttt{description}, and other columns. 
However, the book records it inspects originate from
\ttt{var\_call\_he4GeOzFNhdoBfpZPxkmX09E} (Line 110), 
which is produced by the \ttt{query\_db} call in Line 51 executing
\ttt{SELECT book\_id, title, author, categories, description FROM books\_info;}
This query does not retrieve the \ttt{details} column. 
In effect, 
the agent searches for ``English'' using columns other than the correct one (\ttt{details}), 
leading to failure.

\subsubsection{\fmbadge{FM4}: Incorrect Regular Expression}

\topic{Query}
Q1

\topic{Analysis}
This trajectory fails due to an overly permissive regular expression. 
The agent uses a pattern to extract any four-digit string starting with 19 or 20 (Line 66) 
and then selects the smallest extracted value (Line 73) 
as the publication year. 
This approach may incorrectly extract 
\ttt{1932} from an ISBN 
rather than the true publication year, \ttt{2004}, 
as in the example shown in Line 17.

\end{document}